\documentclass[10pt, prb, superscriptaddress, twocolumn]{revtex4-1}
\usepackage{amsmath}
\usepackage{graphicx}
\usepackage{pdfpages}
\usepackage{pgffor}

\makeatletter
\AtBeginDocument{\let\LS@rot\@undefined}
\makeatother

\graphicspath{{./figs/}}

\newcommand{\FEEaffiliation}{\affiliation{Faculty of Engineering and the Environment, University of Southampton, Southampton, SO17~1BJ, United Kingdom}}
\newcommand{\ie}{\emph{i.e.}}

\begin{document}

\title{Thermal stability and topological protection of skyrmions in nanotracks}

\author{David Cort\'es-Ortu\~no}
\email{d.i.cortes@soton.ac.uk}
\FEEaffiliation

\author{Weiwei Wang}
\FEEaffiliation
\affiliation{Department of Physics, Ningbo University, Ningbo, 315211, China}

\author{Marijan Beg}
\FEEaffiliation

\author{Ryan A. Pepper}
\FEEaffiliation

\author{Marc-Antonio Bisotti}
\FEEaffiliation

\author{Rebecca Carey}
\FEEaffiliation

\author{Mark Vousden}
\FEEaffiliation

\author{Thomas Kluyver}
\FEEaffiliation

\author{Ondrej Hovorka}
\FEEaffiliation

\author{Hans Fangohr}
\email{h.fangohr@soton.ac.uk}
\FEEaffiliation
\affiliation{European XFEL GmbH, Holzkoppel 4, 22869 Schenefeld, Germany}

\begin{abstract}

Magnetic skyrmions are hailed as a potential technology for data storage and
other data processing devices. However, their stability against thermal
fluctuations is an open question that must be answered before skyrmion-based
devices can be designed. In this work, we study paths in the energy landscape
via which the transition between the skyrmion and the uniform state can occur
in interfacial Dzyaloshinskii-Moriya finite-sized systems. We find three
mechanisms the system can take in the process of skyrmion nucleation or
destruction and identify that the transition facilitated by the boundary has a
significantly lower energy barrier than the other energy paths. This clearly
demonstrates the lack of the skyrmion topological protection in finite-sized
magnetic systems. Overall, the energy barriers of the system under
investigation are too small for storage applications at room temperature, but
research into device materials, geometry and design may be able to address
this.

\end{abstract}

\maketitle

The current paradigm of magnetic information storage technology, a technique
called perpendicular media recording, is coming to a limit where magnetic
grains, which are used as information bits, cannot be reduced further in size,
since these units lose stability and information can no longer be recorded. New
methods for further increases in data storage capacity are desired. One of the
main constraints for any data processing and storage technology is the
stability of the data, i.e. the robustness of information carriers against
random thermal fluctuations at operating temperature.

To analyse the stability of a system we can compute the energy barrier between
two of its equilibrium states (that can be used together to represent one bit
of information). This barrier is the energy that thermal fluctuations, or any
other excitation, needs to provide to the system to drive one equilibrium state
to the other. The energy barrier can be used to estimate the average time that
the system can remain in each state and thus provides the time scale over which
information can be stored in the device without corruption.

Magnetic skyrmions are considered an alternative technology to address the
challenges in device engineering~\cite{Nagaosa2013, Fert2013, Sampaio2013,
Iwasaki2013, Wiesendanger2016}. Skyrmions are magnetic structures that can be
associated with topologically charged particles. They have interesting
topological properties that contribute to their stability and they are more
energetically efficient to manipulate~\cite{Nagaosa2013, Fert2013, Sampaio2013,
Iwasaki2013, Wiesendanger2016} than magnetic domain walls. Skyrmions arise in
systems with a broken symmetry due to the Dzyaloshinskii-Moriya Interactions
(DMIs) and can be stabilised using magnetic fields or strong anisotropies in
large samples~\cite{Roesler2011, Sampaio2013, Kiselev2011}. On the other hand,
different works~\cite{Wiesendanger2016, Boulle2016, Beg2015, Sampaio2013,
Rohart2013} have indicated that small confined systems made of DMI materials
are suitable for stabilising skyrmions without the need of an external field.

In the context of possible geometries for the fabrication of a skyrmion based
magnetic device, nanotracks have been suggested and a number of studies have
shown how a series of skyrmions can be stabilised and driven along the sample
by means of weak electric currents~\cite{Fert2013, Iwasaki2013, Schutte2014b,
Nagaosa2013, Purnama2015, Zhang2015b}. In this skyrmion-based racetrack memory
design the information bit 0 or 1 would be encoded by a skyrmion's presence or
absence. An important parameter that needs to be understood is the thermal
stability.  More precisely, we need to explore the energy landscape, find paths
which the system can take in the transition between skyrmion and uniform state,
and identify the one with the lowest energy barrier. Based on the size of this
lowest energy barrier, the lifetime of the binary data can be estimated.

In this study we calculate energy barriers associated with the destruction and
creation of skyrmions in thin ferromagnetic nanotracks with interfacial DMIs.
The methods can be extended to magnetic solids with different DMI mechanisms
and different geometries. Our results are relevant to a variety of interfacial
DMI based materials since we analyse these systems in a range of DMI strengths
and thus skyrmion sizes. The energy barrier calculations is achieved through a
numerical technique called Nudged Elastic Band
Method~\cite{Henkelman2000,Schwartz2002,Bessarab2015} (NEBM), which calculates
minimum energy transitions between equilibrium states in magnetic systems.  We
use an optimised version of the algorithm recently proposed by Bessarab et
al.~\cite{Bessarab2015} since they claim that early versions of the
method~\cite{Dittrich2002,Suess2007,Vogler2013,Thiaville2003} lead to
uncontrolled behaviour of the algorithm. In particular, we are interested on
the minimum energy path between two states, since it gives us the smallest
energy required, \ie{} the energy barrier for a configuration to transition to
another state.

Our results show that there are three main minimum energy transitions for
destroying a skyrmion in a nanotrack within a range of DMI magnitudes. The
lowest energy path is one where the boundaries of the system play a major role,
making evident the lack of topological protection of the skyrmion in a finite
sample. The other two paths with larger energy barriers are a skyrmion collapse
and a skyrmion destruction mediated by a singularity.  We describe and simulate
the system using a (Heisenberg-like) discrete spin model because skyrmion
destruction mechanisms that are not mediated by a boundary are forbidden under
a continuum description of the magnetisation field due to the skyrmion
topology. Hence, estimates of the energy barriers of these energy paths within
the micromagnetic model will depend on the numerical discretisation. 

Although recent works have been published about skyrmionic systems with
interfacial DMI~\cite{Rohart2016a,Lobanov2016}, the key difference is that those
works consider large samples, simulating a skyrmion in an infinite system, thus
they have not observed the effects of the boundaries, which we identify as the
most important route due to the lower energy barrier.

Our discussion will start with a brief introduction to the main concepts of the
NEBM and, consequently, the application of the algorithm to nanotracks of
different DMI strength using an atomistic simulation framework. Accordingly, we
use a hexagonal lattice and we find the three aforementioned skyrmion
destruction mechanisms.

\begin{figure}[tp]
    \centering
    \includegraphics[width=\columnwidth]{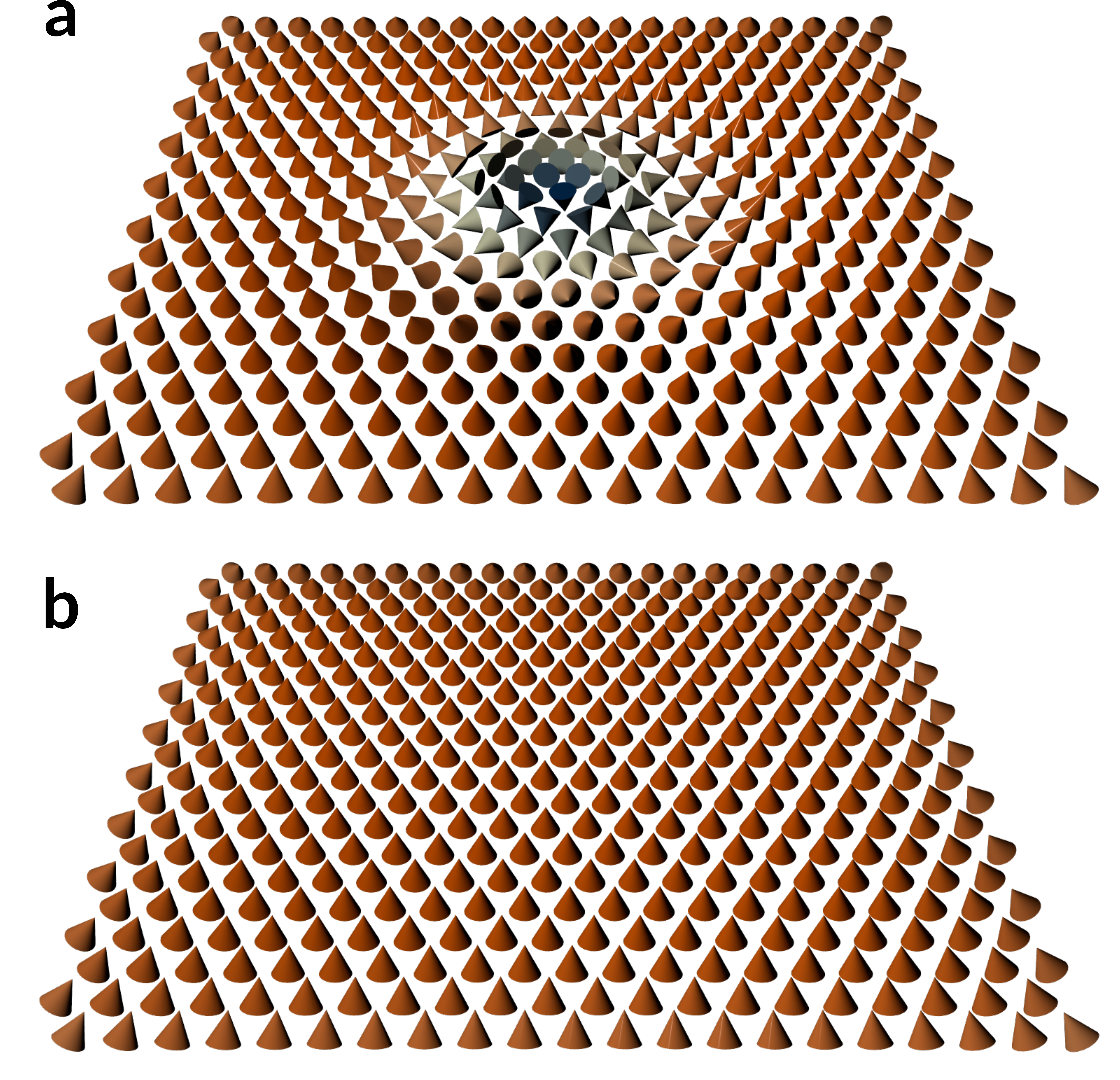}

    \caption{Magnetic configurations in a system with interfacial DMI.
    Representation of: (a) a N\'{e}el skyrmion and (b) a ferromagnetic ordering
    in a thin magnetic film with interfacial DMI.}

    \label{fig:magnetic-states}
\end{figure}%

\section*{Results}

The Nudged Elastic Band Method (NEBM) is an algorithm that searches for minimum
energy transitions between two equilibrium states.  First results of this
numerical method applied to micromagnetics were published by Dittrich et
al.~\cite{Dittrich2002}, computing minimum energy paths (transitions with
minimal energy cost) in a variety of simple magnetic systems, which are
corroborated with analytic theory.  For example, they showed minimum energy
transitions of small particles and elongated particles, where paths are
characterised by coherent rotations and domain wall propagations, respectively.
To help interpret the main results in this work, we need to define some basic
terminology.

\subsection*{Numerical method}

To describe a magnetic material we use a discrete spin model, where we define a
lattice of $P$ nodes which each have associated a three dimensional spin vector
$\mathbf{s}_{i}$, ${i\in\{0,1,\ldots,P-1\}}$. This whole system
$\left(\mathbf{s}_{0},\mathbf{s}_{1},\ldots,\mathbf{s}_{P-1}\right)$ will be
called an \emph{image}, and we will denote it as $\mathbf{Y}$. For example, the
skyrmion and the uniform state shown in Fig.~\ref{fig:magnetic-states} are one
image each. The geometric ordering of the lattice that represents the
arrangement of molecules or atoms, is given by crystallographic nature of the
material. Depending on the magnetic configuration of the magnetic moments, an
image will have a specific energy.  Thus, the energy $E=E(\mathbf{Y})$ of the
magnetic sample is parametrised by the magnetic ordering of $\mathbf{Y}$.

\begin{figure}[tp]
    \centering
    \includegraphics[width=\columnwidth]{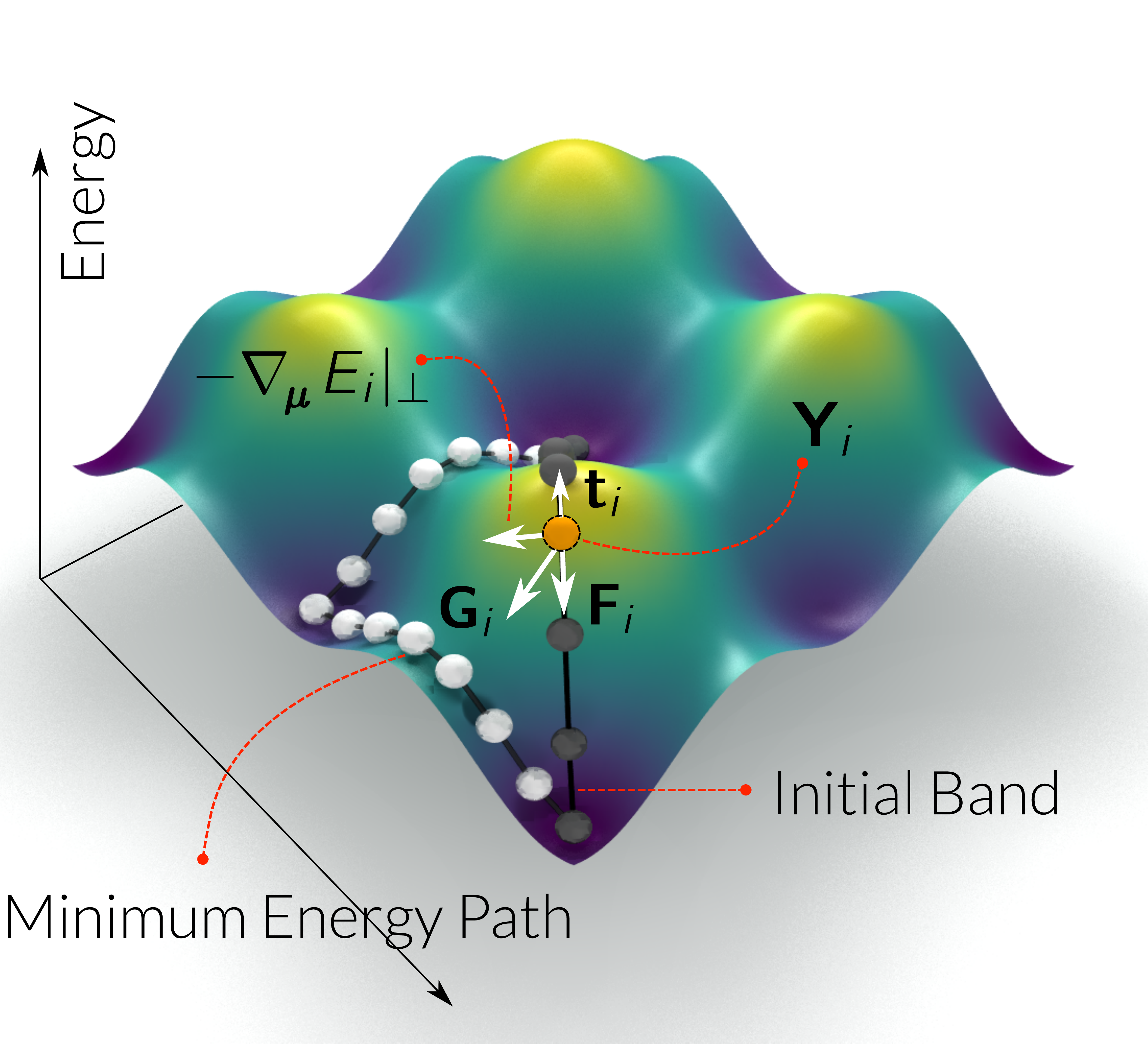}

    \caption{Overview of the NEB method. A summary of the Nudged Elastic Band
        method in a system parametrised by two variables. The surface height
        indicates the energy plotted for all points in this two-dimensional
        phase space. A set of specific magnetic configuration is shown through
        spheres, each being one image of the method. For a particular image
        $\mathbf{Y}_{i}$ of the initial energy band, we show the effective
        (total) force $\mathbf{G}$, the gradient component perpendicular to the
        band $\boldsymbol{\nabla}_{\boldsymbol{\mu}}E|_{\perp}$
        ($\boldsymbol{\mu}=\mu_{\text{s}}\mathbf{s}$ as the magnetic moment, see
        Methods for details), the tangent to the band $\mathbf{t}$ and the
        spring force $\mathbf{F}$. The height of the surface $E_i =
        E(\mathbf{Y}_i)$ shows the energy of image~$i$.  At the extrema of the
        band, the magnetic configurations are fixed and localised at energy minima.}

    \label{fig:nebm_overview}
\end{figure}

In the NEBM, we define a so called \emph{band} of $N$ images $\mathbf{Y}_{i}$,
${i\in\{0,1,\ldots,N-1\}}$, which are identical systems in (ideally) different
magnetic configurations. For each of the images at either end of the band,
$\mathbf{Y}_{0}$ and $\mathbf{Y}_{N-1}$, we fix the magnetic configuration to
be the two equilibrium states for which we want to find the minimum energy
transition.  For the other $N-2$ images,
$\left(\mathbf{Y}_{1},\mathbf{Y}_{2},\ldots,\mathbf{Y}_{N-2}\right)$, we need
to set up an initial sequence of magnetic configurations. A graphical
representation of this set up is shown in Fig.~\ref{fig:nebm_overview} where
the in-plane coordinates represent the two-dimensional phase space, every
sphere $i$ represents a particular magnetic configuration $\mathbf{Y}_i$ in
that phase space, and the surface represents the energy landscape
$E(\mathbf{Y})$. In this figure, the equilibrium states lie in two global
minima and the initial band crosses an energy maximum.

Keeping the images $\mathbf{Y}_{0}$ and $\mathbf{Y}_{N-1}$ fixed, we apply the
NEBM algorithm, which iteratively evolves the band to find the lowest energy
path between these two (see the minimum energy path in
Fig.~\ref{fig:nebm_overview} that passes through a local minimum). This
minimisation is achieved by defining an effective force $\mathbf{G}$ for every
image, which depends on vectors $\mathbf{t}$ that are tangents to the energy
band.  Additionally, since the energy of the band is minimised, it is necessary
to set a spring force $\mathbf{F}$ between the images in order to keep them
equally spaced in the phase space and avoid that the images cluster around the
fixed states.  Accordingly, to distinguish them, we use a Geodesic
\emph{distance}~\cite{Bessarab2015} which measures the difference of the spins
direction between consecutive images (see Methods section).

When the NEBM reaches convergence (see Methods section), the band will ideally
pass through a maximum in energy along a single direction in phase space, which
is known as a first order saddle point and it determines the energy barrier
between the two fixed configurations. This transition path might not be unique
and if it is the one with the smallest energy barrier, we call it the minimum
energy path. For the final energy band shown in Fig.~\ref{fig:nebm_overview},
there are two maxima since the band crosses a metastable state that could be
used as an equilibrium configuration, but we can clearly distinguish a single
saddle point between every pair of energy minima.  In that case the most
relevant first order saddle point would be the one with largest energy, which
is the barrier that the system needs to climb to get to the other equilibrium
state. In general, there is no guarantee that any of the images in the band
sits exactly at the saddle point (commonly, there will be images to either side
of the saddle point along the band), and thus the energy at the saddle point
(and hence the energy barrier) is generally underestimated.  To address this
problem and improve the accuracy of the estimate, we can push one of the images
into that maximum energy position along the path using a variation of the
method called the Climbing Image NEBM~\cite{Bessarab2015,Henkelman2000a}.  This
is based on taking the largest energy point from a relaxed band (with the
NEBM), redefine the forces applied to this image and then remove the spring
force on it (see Methods for details).  As a result, this image will try to
climb up in energy along the band (while being allowed to decrease its energy
in a direction perpendicular to the band).

The NEBM algorithm implemented in our software~\cite{Fidimag2} has been tested
to reproduce basic models proposed by Dittrich et al.~\cite{Dittrich2002},
giving successful results. In addition, we reproduced the skyrmion annihilation
test problem described by Bessarab et al.~\cite{Bessarab2015,Cortes2016a} using
an atomistic model for a two dimensional square lattice of atoms with
interfacial DMI, which is discussed in the Test Model section of the
Supplementary Information.

To explore the suitability of chiral structures for information recording, we
have chosen cobalt samples with DMI for our study, based on recent works on
these systems~\cite{Sampaio2013,Moreau-Luchaire2016,Boulle2016}. This will also
show us another perspective of the claimed topological protection properties of
skyrmions in finite systems: in confined geometries the boundaries play a major
role on the skyrmion stability.  Accordingly,  based on
[\onlinecite{Sampaio2013}], we define an 80 nm long, 40 nm wide and 0.4 nm
thick stripe, which we discretise into a 320 by 185 spin lattice with a lattice
constant of $2.5\text{\AA}$ (see Methods for details). This system has an
interfacial DMI whose magnitude we vary and a strong uniaxial out of plane
anisotropy. The DMI in a Co based system can be obtained by stacking the cobalt
on top of a heavy metal with a larger spin orbit coupling and experimental
techniques have been proposed to tune the DMI
magnitude~\cite{Moreau-Luchaire2016,Ma2016}.  At the time of publication of
Ref.~\onlinecite{Sampaio2013}, there was no experimental evidence of the Co
samples under study, thus the magnetic parameters are based on standard Co
material.  Correspondingly, the atomic layer spacing is assumed as
$a_{z}=2.5\,\text{\AA}$ and a lattice constant of $a=2.5\, \text{\AA}$.  The
atomic arrangement of an FCC cobalt layer has an hexagonal
structure~\cite{Yang2015a,Sampaio2013,Rohart2016a}.

\begin{figure}[tp]
    \centering
    \includegraphics[width=0.95\columnwidth]{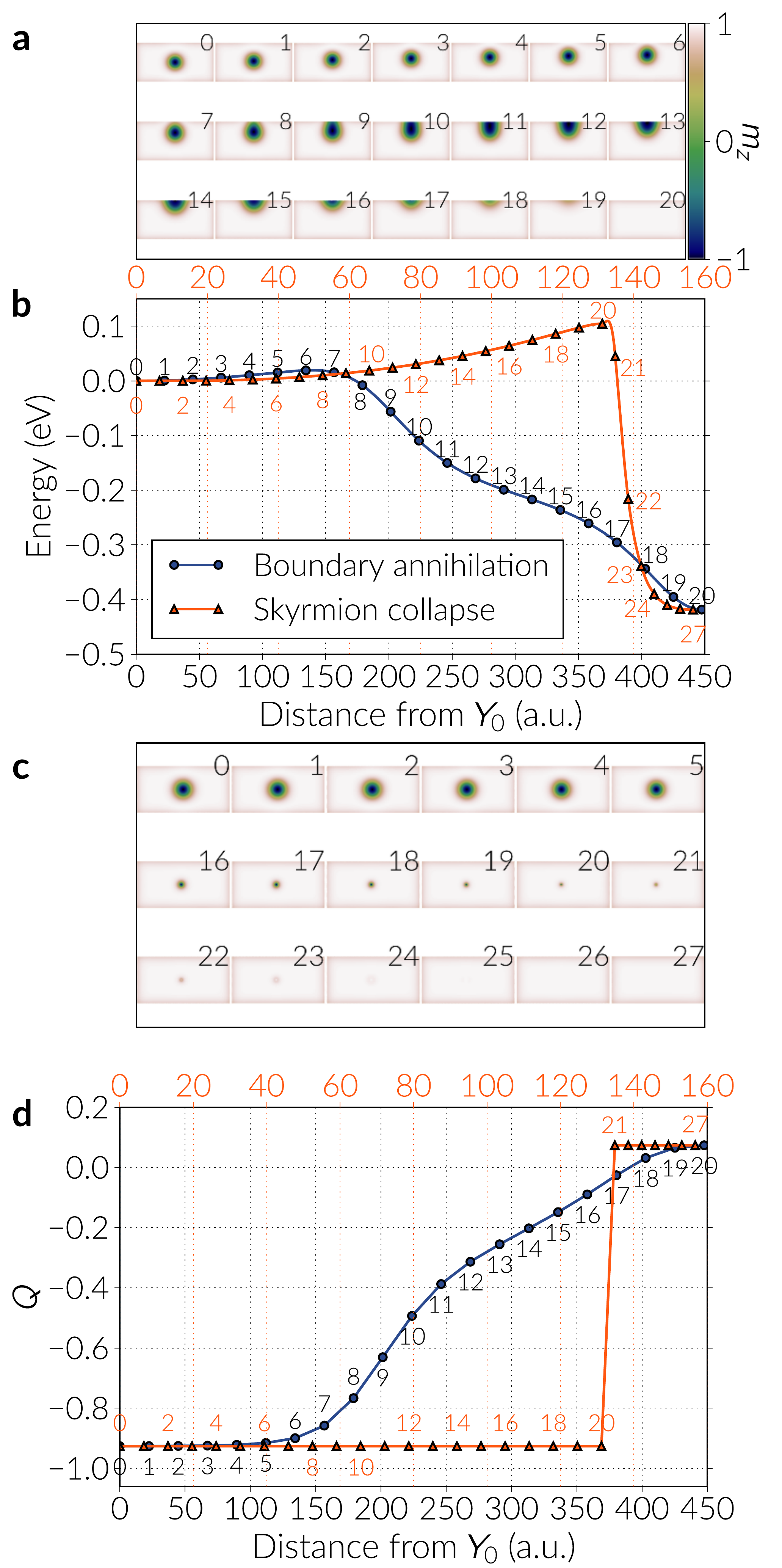}

    \caption{Minimum energy paths of a skyrmion in a cobalt nanotrack. The DMI
    constant of the system is $D=0.676\,\text{meV}$ of magnitude.  There are
    two different paths: a skyrmion annihilation at a boundary and a
    symmetrical skyrmion collapse. (a) Images of the band for the boundary
    annihilation, annotated according to the numbers in the corresponding curve
    in (b). The colour scale refers to the out of plane ($z$) component of the
    magnetisation field. (b) Energy bands for both minimum energy paths as a
    function of the distance from the first image (left extreme of the bands).
    The top scale refers to the skyrmion collapse case. (c) Images of the band
    for the skyrmion collapse. (d) Topological charge (skyrmion number) $Q$ as
    a function of the images distances for the cases depicted in (b). The top
    scale refers to the skyrmion collapse case. }

    \label{fig:nanotrack-transitions-D3}
\end{figure}%

Including the strong anisotropy and the confined geometry, the interfacial DMI
in a Co layer favours the stabilisation of chiral structures. In general, we
observe four well defined equilibrium states, within the range of DMI
magnitudes we study. Two of them are N\'{e}el skyrmions with the core pointing
up or down with respect to the out of plane ($z$) direction (see
Fig.~\ref{fig:magnetic-states}a) and that are degenerate in energy.  Similarly,
the two other are degenerate ferromagnetic orderings pointing perpendicular to
the nanotrack plane (Fig.~\ref{fig:magnetic-states}b), with a small canting at
the boundary of the system. We obtained these states by relaxing similar
configurations using the Landau-Lifshitz-Gilbert equation.  However, we cannot
guarantee that these are the only equilibrium configurations since, depending
on the $D$ value, other chiral orderings can arise but they cannot be easily
identified.  Generally, knowing the true global minima of a specific system is
not straightforward, but we will focus on the four aforementioned magnetic
states, using them as fixed NEBM images at the extrema of the energy bands.

\subsection*{Nanotracks}

We begin our analysis by performing a systematic study of long tracks with
different DMI constants. We define the DMI in a range from 2.6 up to $3.6
\,\text{mJ}\,\text{m}^{-2}$ in steps of $0.2\,\text{mJ}\,\text{m}^{-2}$.  To
apply these values to a discrete spin model, we converted them from
micromagnetic values considering the hexagonal nature of the sample.
Therefore, the equivalent atomistic DMI constants in the discrete model range
from $D=0.586\,\text{meV}$ to $0.811\,\text{meV}$ in steps of
$0.045\,\text{meV}$.

In the range of DMI magnitudes we choose, the skyrmion energy is always larger
than the uniform configuration energy, but it starts to get closer to that of
the ferromagnetic state as the $D$ value increases.  Additionally, the systems
differ in skyrmion size, where a skyrmion gets larger as the DMI constant
increases (see Supplementary Fig.~S2 for detailed values of the skyrmion
sizes). The length of the track is not relevant as long as the skyrmion size is
not affected by the long edge boundaries, since the isolated skyrmion does not
interact with them. On the other hand, the stripe width is defined according to
Ref.~\onlinecite{Sampaio2013}, which is reasonable for a novel technological
application, and the skyrmion will slightly interact with the short edge
boundaries when the DMI is strong enough. Larger DMI magnitudes than the values
we specify are not analysed since skyrmions acquire an elongated
shape~\cite{Sampaio2013,Yoo2014} and the skyrmion loses its symmetrical
character. A different method to modify the skyrmion dimensions, rather than
varying the DMI, is tuning the uniaxial anisotropy, where a stronger anisotropy
reduces the skyrmion size. However its mechanism is different to the
antisymmetric exchange, thus the energy landscape is likely to change and
analysing this effect goes beyond the scope of this study.

For every case, we use two different initial bands for the NEBM: (i) a linear
interpolation using spherical coordinates, which means interpolating the
spherical angles that describe the magnetic moments and (ii) a skyrmion
displacement towards one edge of the disk, making it disappear at the boundary.

After relaxation with the NEBM, we obtained three different transitions.  One
of these paths is a symmetric skyrmion collapse (shrinking) until the last spin
at the skyrmion centre flips to give rise to the ferromagnetic ordering, which
originates from the linear interpolation initial state.  The second transition
is given by the annihilation of the skyrmion core through a singularity that
resembles a Bloch point, and this is observed for DMI magnitudes of
$0.676\,\text{meV}$ and above.  The third transition we observe is given by the
displacement of the skyrmion towards the boundary, where the skyrmion
configuration is deformed until annihilation. Detailed images of these three
different transitions and the initial states are provided in Supplementary
Fig.~S3 and S4.

\subsection*{Boundary annihilation}

We observe the skyrmion annihilation at a boundary for every DMI value. In
particular, for the case of $D=0.676\,\text{meV}$, we show in
Fig.~\ref{fig:nanotrack-transitions-D3}b the energy band for this path and in
Fig.~\ref{fig:nanotrack-transitions-D3}a, the corresponding images where the
skyrmion is annihilated at the boundary (see also Supplementary Video~S4).
Additionally, in Fig.~\ref{fig:nanotrack-transitions-D3}d, we illustrate the
total topological charge of the images where a smooth transition towards the
uniform state occurs.

\begin{figure}[tp]
    \centering
    \includegraphics[scale=0.4]{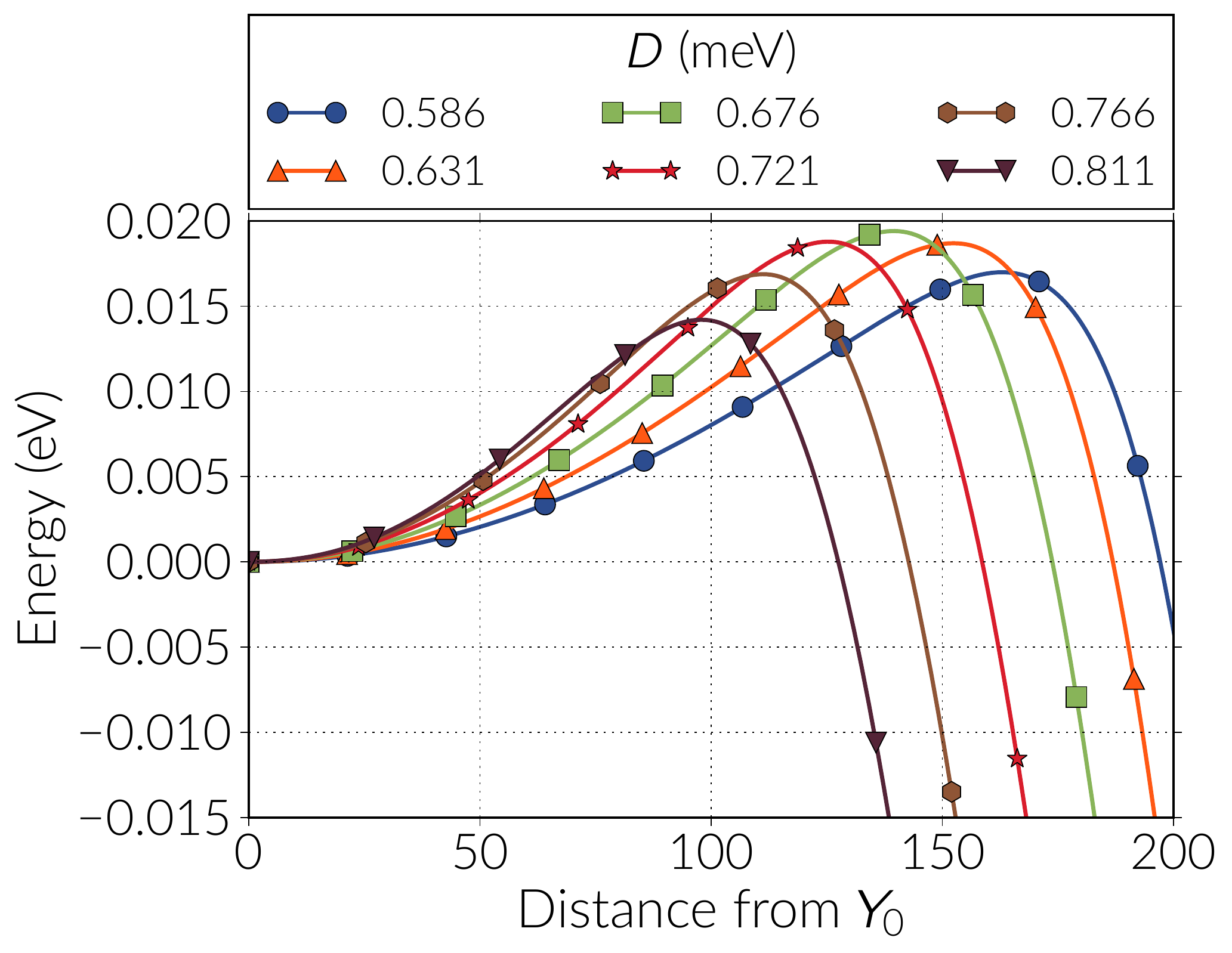}

    \caption{Energy bands for the skyrmion annihilation at the boundary. The
    bands are shown for different DMI constant $D$ and only for the first few
    images of the bands. The first or extreme left image is the skyrmion
    configuration and, for every case, the energy is redefined with respect to
    the skyrmion energy. The continuous     line is a cubic polynomial
    interpolation.}

    \label{fig:nanotrack-bands-ba-D}
\end{figure}%

\begin{figure}[tp]
    \centering
    \includegraphics[width=0.88\columnwidth]{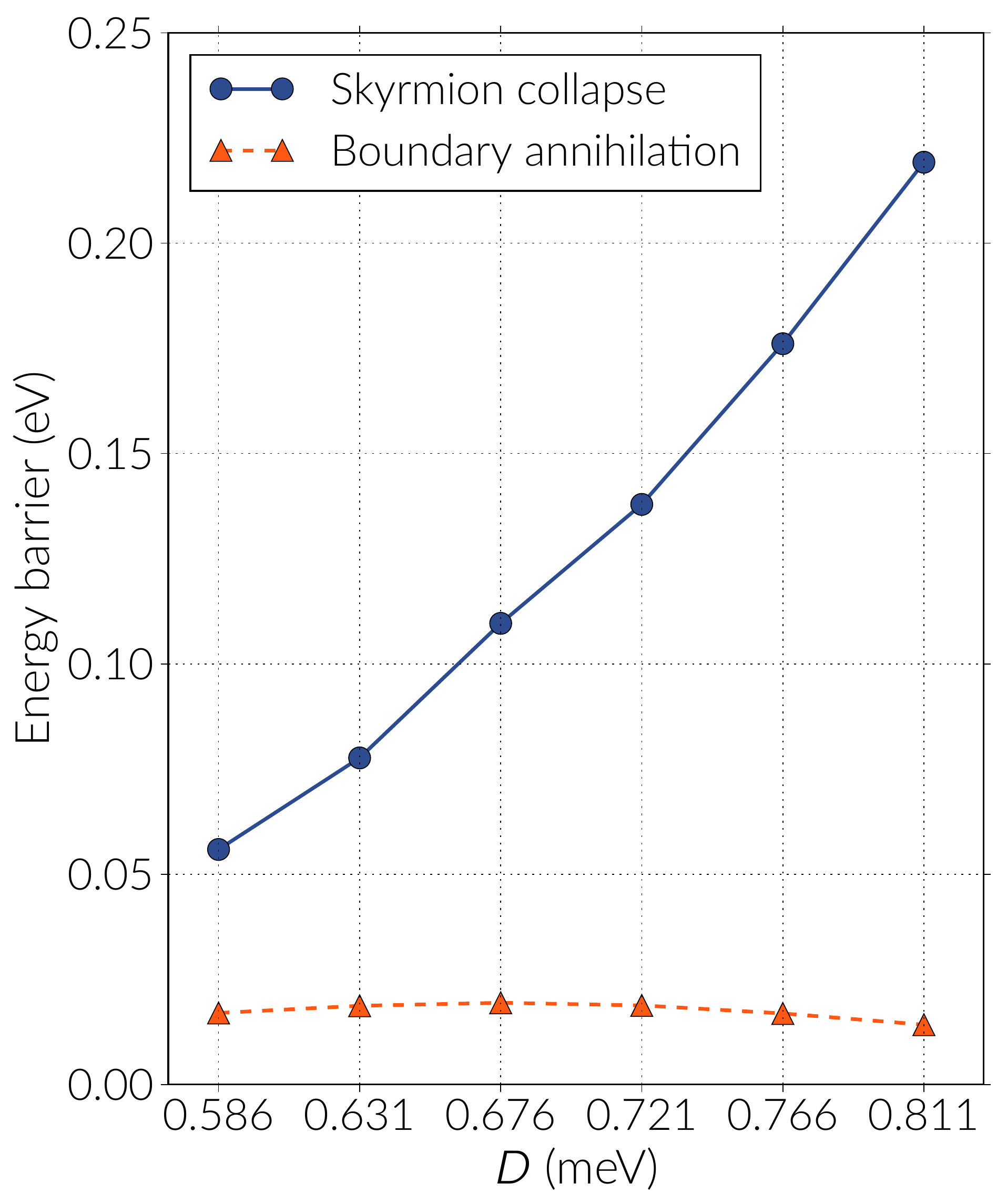}

    \caption{Energy barriers for two different energy paths. The energy barrier
        dependence on the DMI constant is shown for skyrmion collapse and
        skyrmion annihilation at the boundary.  The energy barriers are
        calculated with respect to the skyrmion energy and using a cubic
        polynomial interpolation on the images of the energy bands.}

    \label{fig:nanotrack-barriers-D}
\end{figure}

We analyse the dependence of the energy bands with respect to the DMI strength
in Fig.~\ref{fig:nanotrack-bands-ba-D} for the first few images of every band,
and in Fig.~\ref{fig:nanotrack-barriers-D} we plot the energy barriers with
respect to the skyrmion energy using the continuous curve approximation, which
is a cubic polynomial that uses information from the tangents of the images in
the band (a comparison of the energy barriers values using the approximation
with respect to the actual data points is shown in the Supplementary
Information). The tendency in Fig.~\ref{fig:nanotrack-barriers-D} is that the
barriers decrease almost quadratically with larger DMI magnitudes, where a
maximum is observed around $D=0.676\,\text{meV}$ (see
Table~\ref{tab:energy-barriers}). In general, these barriers are significantly
smaller than the energy difference between the skyrmion and the ferromagnetic
state. To check the robustness of the results we also modified the NEBM spring
constant values and we found out that there are small variations of the
barriers when changing this parameter, which is mostly due to the image
positioning in the band, and if the spring constant is too large the method
struggles to converge.  In the results of Fig.~\ref{fig:nanotrack-bands-ba-D},
we can notice that when increasing the DMI strength, the resolution of the
images before the saddle point gets poorer, however these variations are not
large and applying the climbing image method to the largest energy states gives
energy barrier magnitudes similar to those obtained when performing a
polynomial approximation of the band.

In the range of DMI magnitudes we analysed the skyrmion destruction at the
boundary is the transition that has the lowest activation energy.
Specifically, the energy barriers are an order of magnitude smaller than the
skyrmion collapse transitions. This demonstrates the lack of topological
protection for the skyrmions, which is due to the finiteness of the system.

\subsection*{Skyrmion destruction}

When relaxing energy paths that directly involve the destruction of the
skyrmion core in a region inside the track, the spins at the center of the
skyrmion reverse to form a ferromagnetic state. This is a non trivial process
since the band suffers drastic changes in energy when the NEBM tries to
converge to a minimum energy transition and the system must undergo a
topological change.  This path is (according to our observations) the most
likely for a skyrmion situated in a large or infinite sample, where it is meant
to be topologically protected, thus we expect a larger energy barrier than in
the transition mediated by a boundary.

\begin{figure}[tp]
    \centering
    \includegraphics[width=\columnwidth]{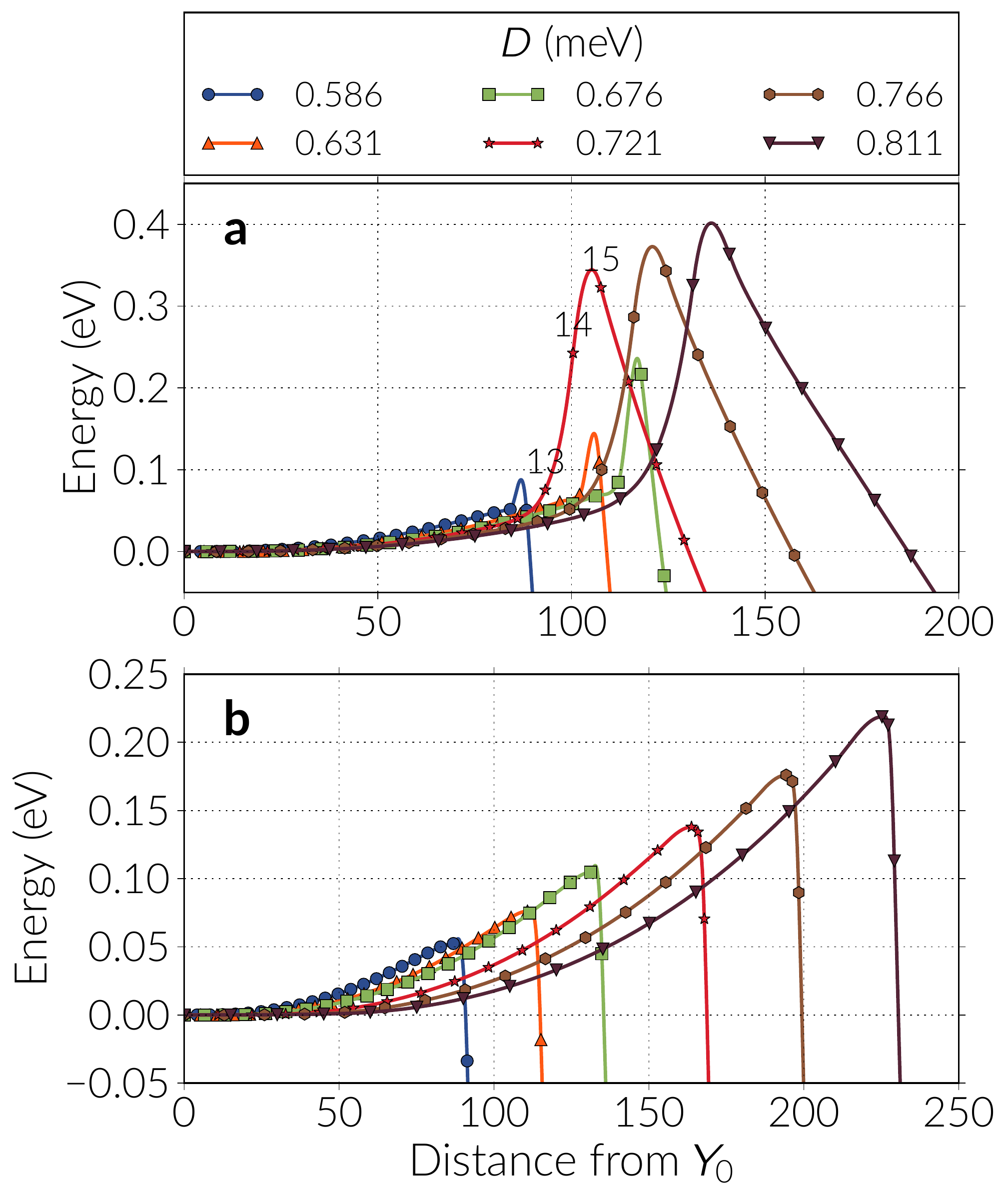}

    \caption{Energy bands for two skyrmion destruction mechanisms. The bands
    are shown for different DMI constant $D$ and only for the first few images
    of the bands. The first or extreme left image is the skyrmion configuration
    and for every case, the energy is redefined with respect to the skyrmion
    energy. The continuous line is a cubic polynomial interpolation. (a) The
    energy bands obtained with the NEBM.  The curves for DMI values of
    $D=0.676\,\text{meV}$ and below, are the skyrmion collapse. For values of
    $D=0.721\,\text{meV}$ and above, the bands are the skyrmion destroyed by a
    Bloch point like singularity. (b) Refined energy bands obtained with the
    Climbing Image NEBM applied to the largest energy points of the curves in
    Figure (a). We expect the data in (b) to be a better approximation of the
    energy barrier. For this case, all the bands converged towards the skyrmion
    collapse path.}

    \label{fig:nanotrack-skcollapse-bands-D}
\end{figure}

We firstly found that for DMI magnitudes of $D=0.676\,\text{meV}$ and below,
the algorithm converged to the skyrmion collapse process, which is similar to
the one depicted in Fig.~\ref{fig:nanotrack-transitions-D3}c. In
Fig.~\ref{fig:nanotrack-skcollapse-bands-D}a we show the first images of the
bands, where the skyrmion state is given by the left extrema, and we observe
pronounced peaks at the saddle points. These points are the images that have a
tiny skyrmion with only a few spins defining its core before reversing.  It is
worth noting that these saddle points have a finite energy since we have a
discrete number of magnetic moments, whereas in a continuum model it is likely
that the peak depends on the discretisation of the continuum mesh that defines
the material. In our results, the saddle point energies (and thus the energy
barriers) increase with the DMI constant, where values range between 0.1 and
0.25~eV larger than the skyrmion energy. Around the saddle points, the energy
landscape must have a rough shape since the neighbouring images usually have a
substantial change in energy.  Specifically, we observed that when the images
move along the band before the algorithm reaches convergence, the images that
cross this region suffer large energy alterations (see Supplementary Video~S1
as an example). In order to resolve more accurately the energy value of the
saddle points, we applied the climbing image technique on them, slightly
reducing their energy and improving the resolution around them. We show these
results in Fig.~\ref{fig:nanotrack-skcollapse-bands-D}b (see also Supplementary
Video~S5).  The value of the energy barriers are summarised in
Fig.~\ref{fig:nanotrack-barriers-D} and Table~\ref{tab:energy-barriers}, where
the magnitudes increase almost linearly with the DMI magnitudes. Furthermore,
as we did for the boundary annihilation, using the case of
$D=0.676\,\text{meV}$, we plot in Fig.~\ref{fig:nanotrack-transitions-D3}b the
energy band of the skyrmion collapse path and the corresponding images in
Fig.~\ref{fig:nanotrack-transitions-D3}c. We notice that the saddle point lies
between the 20th and the 21st image, where the last few spins at the tiny
skyrmion core reverse, and there is a drastic change in energy in the next
images. A different perspective of this phenomenon is observed when looking at
the topological charge values of the images in
Fig.~\ref{fig:nanotrack-transitions-D3}d. Up to the 20th image, the images
preserve the skyrmion structure, thus having equivalent skyrmion number.
However, after this point the skyrmion core is reversed and the skyrmion
ordering is lost, which is indicated by a sharp change in the topological
charge magnitude.

\begin{table}[tp]
    \def\arraystretch{1}%
    \begin{tabular}{l r r r}
        \hline
                  & \multicolumn{3}{c}{Energy barrier (eV)} \\
                  \cline{2-4}
        $D$ (meV) & Boundary & Collapse & Singularity \\
        \hline
        0.586     & 0.0170   & 0.0559   &         \\
        0.631     & 0.0187   & 0.0776   &         \\
        0.676     & 0.0194   & 0.1096   &         \\
        0.721     & 0.0188   & 0.1379   &  0.3445 \\
        0.766     & 0.0169   & 0.1761   &  0.3728 \\
        0.811     & 0.0142   & 0.2193   &  0.4016 \\
        \hline
    \end{tabular}
    
    \caption{Energy barriers for the skyrmion destruction energy paths. Energy
        values are computed with respect to the skyrmion energy for different
        DMI constants. The singularity driven skyrmion destruction is only
        observed for strong enough DMI constant magnitudes.  }
    
    \label{tab:energy-barriers}
\end{table}

\begin{figure*}[tp]
    \centering
    \includegraphics[width=\textwidth]{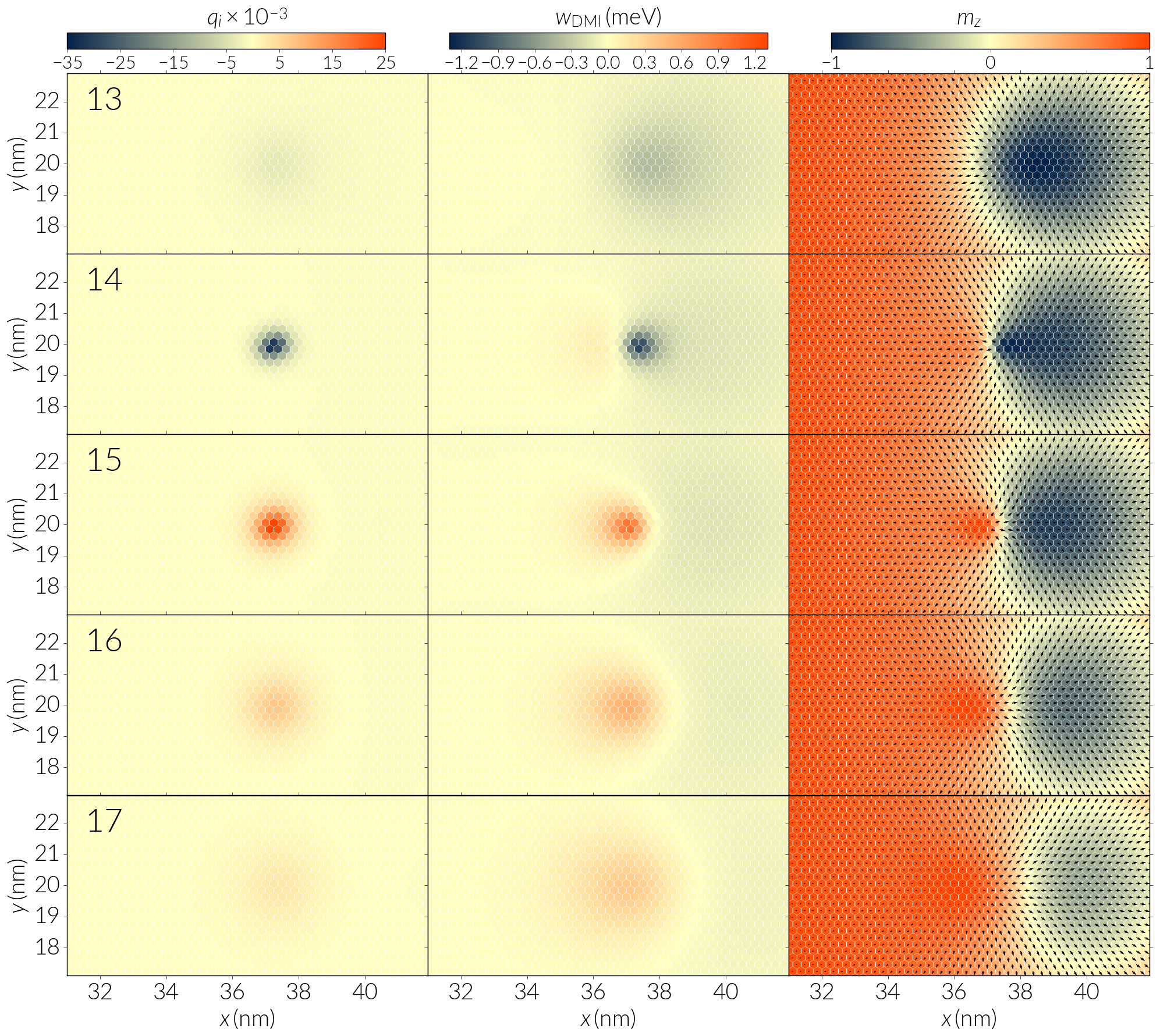}

    \caption{Skyrmion destruction mediated by a topological singularity. The
    sequence shows the snapshots of the energy band images for an nanotrack
    system of DMI constant $D=0.721\,\text{meV}$, described by a hexagonally
    arranged discrete spin lattice.  This band corresponds to the annotated
    curve of Fig.~\ref{fig:nanotrack-skcollapse-bands-D}a.  The left column
    shows the topological charge density, the middle sequence the DMI energy
    density per lattice site and the right column the corresponding
    magnetisation fields.  The hexagonal mesh can be distinguished from the
    honeycomb shaped data points.  The numbers at the top left of every row
    indicate the image number from the energy band.}

    \label{fig:nanotrack-D32-singularity}
\end{figure*}%

For DMI magnitudes of $D=0.766\,\text{meV}$ and above, the NEBM converged to a
path where the skyrmion core is destroyed by the emergence of a singularity
close to the skyrmion boundary. These paths have associated energy barriers of
above $0.3\,\text{eV}$ (see Table~\ref{tab:energy-barriers}) and we show the
first images of the bands in Fig.~\ref{fig:nanotrack-skcollapse-bands-D}a (see
Supplementary Video~S2 for the evolution of the band and Supplementary Video~S6
for the transition). The singularity that drives the skyrmion towards the
ferromagnetic state resembles a Bloch point, and
Fig.~\ref{fig:nanotrack-D32-singularity} shows the sequence of images of the
band where this structure destroys the skyrmion, for the system with a DMI
constant of $D=0.721\,\text{meV}$. This series of snapshots are zoomed around
the skyrmion core and the numbers at the top left of every row indicate the
image numbers in the band, which correspond to the annotated points of the
curve with stars in Fig.~\ref{fig:nanotrack-skcollapse-bands-D}a. The left
column in Fig.~\ref{fig:nanotrack-D32-singularity} is the topological charge
per lattice site $q_{i}$, the middle column is the DMI energy density and the
right column shows the spin vector field where the colors indicate the
component of the magnetisation perpendicular to the track plane, $m_{z}$. At
the 14th image of the sequence (second row), we observe that the skyrmion core
concentrates to the left side of the original skyrmion core with a drop in
negative DMI energy.  Moreover, the topological number is reduced since the
spins in a tiny region cover most of the directions in a unit sphere, like a
small skyrmion (at the 13th snapshot the topological charges are smaller since
the different spin directions are more spread out). Consequently, at the 15th
image a singularity emerges, which has a chirality opposite to the skyrmion,
indicated by the positive gain in DMI energy and a positive charge. At this
point, the skyrmion core has already been annihilated. The singularity has a
hedgehog structure occupying a radius of around 3 lattice sites, and it will
expand to give rise to the ferromagnetic state at the end of the energy band,
decreasing the DMI energy. Characterising this singularity as a Bloch point is
not trivial due to its two dimensional nature and the lacking of an appropriate
resolution to mathematically define it. In fact, the critical process lies
between the 14th and 15th image and if we refer at
Fig.~\ref{fig:nanotrack-skcollapse-bands-D}a, the 15th image is close to the
saddle point of the band. Interestingly, diverse
studies~\cite{Sampaio2013,Elias2014,Yin2016} have reported the mechanism of
destroying or nucleating a skyrmion by means of a non trivial topological
structure, when applying spin polarised currents. In particular, El\'{i}as and
Verga~\cite{Elias2014} characterise this singularity by its internal magnetic
field, which relates to the topological charge, and Sampaio et
al.~\cite{Sampaio2013}  show how a singularity concentrates DMI energy at the
border of the skyrmion before it is nucleated (see Fig.~S3 of their
Supplementary Information). This transition is not a minimum energy path but it
is an alternative path to disrupt the skyrmion stability that seems to be
preferred by the system when following certain dynamical processes.

Although the bands where a singularity destroys the skyrmion converged without
exhibiting the energy fluctuations observed in the cases with smaller DMI
constants (see Supplementary Fig.~S5), they are not very stable and are likely
to pass through a higher order saddle point.  This is because after applying
the climbing technique on the images at the saddle points, the bands converged
towards the skyrmion collapse transitions. These paths have smaller energy
barriers, as shown in Fig.~\ref{fig:nanotrack-skcollapse-bands-D}b, and the
energy barrier has an approximately linear dependence on the DMI magnitude, as
mentioned previously.

Regarding the simulations with a climbing image, it is usually the case that
this variation of the algorithm helps the saddle point to climb up in energy
along the band and thus improve the value of the energy
barriers~\cite{Bessarab2015,Henkelman2000a}. However, for the skyrmion
simulations, the largest energy points decreased their energy, as seen in
Fig.~\ref{fig:nanotrack-skcollapse-bands-D}. Furthermore, we analysed the
evolution in energy of these saddle points with respect to the number of
iterations of the algorithm, and we saw that they oscillate for a while before
reaching a more stable state (see Supplementary Fig.~S5 and Supplementary
Video~S3), which is given by the symmetrical destruction of the skyrmion, and
these oscillations are prolonged for systems with larger DMI magnitudes. A
plausible explanation for this phenomenon is as follows. As we mentioned
before, the energy landscape has a rough shape or small peaks in energy around
a critical point. In particular, this must be caused because a singularity that
destroys a skyrmion could appear in any place around the skyrmion boundary,
giving multiple possibilities for a saddle point.  Additionally, besides a
component of the effective force that makes a climbing image to go up in energy
along the band, there is also a component that still allows it to follow a
direction of minimum energy.  Therefore, as the image tries to climb, it may
also be pushed to a small minimum energy region if it is over a very narrow
peak in the landscape, decreasing its energy.  Through these dynamics, the
climbing image can, overall, move to a region of lower energy and possibly to a
smoother energy region, where it finally settles. Thus, according to our
results, the climbing image is a mechanism that helps us to find more
equilibrium solutions for the minimum energy transitions.

\section*{Discussion}

Studying the paths through phase space along which a skyrmion can be
annihilated from a track or created in the track, we find that for the geometry
studied the lowest energy barrier is generally found by using the boundary of
the track (see transition snapshots in
Fig.~\ref{fig:nanotrack-transitions-D3}a).  This transition path circumvents
the topological protection, as has been pointed out by Streubel et
al.~\cite{Streubel2015} explaining that the skyrmion topological protection is
non-existent due to the system finiteness.  Therefore, the low stability of
these configurations found by the NEBM simulations seems reasonable.  We
quantify this stability by analysing the results using the Arrhenius-N\'eel law
for the relaxation time

\begin{equation}
    \tau = \tau_{0} \exp\left(\frac{\Delta E}{k_{\text{B}} T}\right)
    \label{eq:arrhenius-neel}
\end{equation}

\noindent where $\tau_{0}$ is related to the attempt frequency $f_{0} =
\tau_{0}^{-1}$, $T$ is the temperature and $k_{\text{B}}$ is the Boltzmann
constant.  The attempt frequency magnitude is not easy to obtain since the
theory normally refers to simple or macrospin systems, but its value is usually
in the range~\cite{Schrefl2001,Aharoni2000} $10^{9}$ to $10^{12}$~Hz.  Using
our computed values for the skyrmion energy barriers and
$f_{0}=10^{9}\,\text{Hz}$, we obtain, at room temperature: (i) for the boundary
annihilation in nanotracks with $D=0.676\,\text{meV}$ (barrier of approximately
0.0194 eV) an average lifetime of $\tau=2.117\,\text{ns}$, whereas (ii) for the
skyrmion collapse, the barrier is 0.110 eV, thus this gives us
$\tau=70.347\,\text{ns}$.  We must notice that the exponential is very
sensitive to the energy barrier values, which can make a difference if the
system follows the skyrmion collapse path, and $f_{0}$ could be smaller, which
would make the lifetimes some orders of magnitudes larger.  Although these
times seem very small, they are computed at room temperature, hence an
experiment~\cite{Romming2013} at 4.2~K, for example, would make a relaxation
time of $\tau=1.864\times 10^{14}\,\text{s}$. To confirm the phenomenon of low
barriers, it would be ideal to compare experimental data showing skyrmion
switching times that could be used to estimate an attempt frequency and
exponential law for its average lifetime.

Moreover, we can compare our calculated energy barriers with respect to recent
magnetic recording technology. In the context of heat assisted magnetic
recording technology, using the data from Weller et al.~\cite{Weller2014} and
approximating a magnetic grain with a spherical shape, the energy barrier for
that system is approximately $95.54\,k_{\text{B}}T$ at room temperature
($T=300\,\text{K}$). In addition, from typical parameters of magnetic grains in
perpendicular recording media, provided by Richter~\cite{Richter2007}, we can
estimate energy barriers of about $97.06\,k_{\text{B}}T$. On the other hand,
for the skyrmionic system of $D=0.676\,\text{meV}$, the boundary annihilation
gives us a barrier of only $0.75\,k_{\text{B}}T$ and the skyrmion collapse a
barrier of $4.24\,k_{\text{B}}T$, which substantiates our findings of low
thermal stability of skyrmions in confined systems.

This phenomenon of low stability must also be present in other finite
geometries, like cylindrical structures that have been proposed
recently~\cite{Beg2015,Rohart2013,Sampaio2013}, complicating any technological
application of these systems, at least at room temperature, since at lower
temperatures the energy barrier can be significant.  For skyrmion based
technology to operate at room temperatures, we need to find systems with larger
energy barriers for skyrmion destruction; other DMI hosting materials such as
those with a bulk interaction, or those of increased thickness may have larger
barriers. This claim is only speculative since modifying the parameters
involved in the system, such as adding thickness to the sample or applying
external magnetic fields, changes the energy landscape and different paths
would be encountered with the NEBM relaxation, which would require a proper
analysis in a different study. On the other hand, a recent theoretical
study~\cite{Muller2016} has proposed taking advantage of edge instabilities for
the creation of skyrmions through the boundaries of a track, avoiding other
larger topological energy barriers from different energy paths.

The processes of destroying the skyrmion by a collapse or by a singularity, are
energetically feasible due to the discrete nature of the crystal lattice. In
the case of an infinite system (i.e. no geometry boundaries), we have observed
so far that these paths are the only possible ways to destroy a skyrmion. Thus
having a skyrmion far away from boundaries is an option to get more stability
but requires larger structures. Regarding the Bloch point like transition, we
mentioned that it this not observed for the cases with weaker DMIs.  We believe
this is because it might be difficult to resolve this structure, which spans a
circle of about 3 lattice points of diameter, when the skyrmion is sufficiently
small. Our preliminary results on samples of Fe on Ir, where skyrmions are only
a few nanometres wide, have confirmed this. 

We must emphasize that we are using an atomistic spin model, where each spin
represents one atomic magnetic moment and which requires more computational
effort than a micromagnetic model. However, a continuum description of the
magnetisation field is inaccurate since the skyrmion collapse or Bloch point
occurring in the higher energy paths we find, break the assumption of
micromagnetics that the magnetisation changes slowly as a function of position.
Because of this violation, it is not possible to predict quantitatively the
energy associated to the saddle points of these paths, giving the energy
barriers dependence on the sample discretisation.

Furthermore, for the skyrmion collapse path the energy barriers are larger for
increasing DMI values. This is because the energy of helicoidal configurations
approaches that of the ferromagnetic state, reaching a point where helicoids
become the ground state of the system. Additionally, larger DMI magnitudes
favour multiple twistings of the magnetisation. Therefore, paths that directly
involve helicoidal structures are significantly preferred. This is also the
reason why the boundary annihilation has very low barriers for strong DMI
cases.

For DMI magnitudes below the range we have studied, the energy barriers become
smaller and we noticed that at $D=0.450\,\text{meV}$ the skyrmion collapse has
a slightly smaller energy barrier than the boundary annihilation, which is
expected from the tendency of the curves of
Fig.~\ref{fig:nanotrack-barriers-D}.  However, this DMI value is close to a
critical value of $D=0.405\,\text{meV}$, below which a skyrmion cannot be
stabilised anymore. These results are shown in the Supplementary Information.

A recent publication~\cite{Rohart2016a} reports a study on cobalt monolayers by
means of the NEBM and two different observed transition paths.  Their Path~1 is
similar to the skyrmion collapse we have reported in this work and, recently,
by Lobanov et al.~\cite{Lobanov2016}. Their Path~2 adds a rotation of the spins
before the core collapses, which resembles a horizontal cut of a three
dimensional Bloch point.  Although we observe this topological structure, in
our case it usually appears at the skyrmion core boundary for large enough
skyrmion sizes, which depend on the DMI magnitude.  We have only observed that
a skyrmion collapse with rotating spins appears as the initial state when we
set up the initial band using linear interpolations, but which disappear as the
elastic band reduces its energy. This has also been confirmed by Lobanov et
al.~\cite{Lobanov2016}. In this context,  a key difference between this
skyrmion stability study and Ref.~\onlinecite{Lobanov2016,Rohart2016a} is that
we take the role of the boundary into account. We find that skyrmion
annihilation and creation via the boundary has an order of magnitude lower
energy barrier for this track geometry and will thus be the preferred path for
the system.

In summary, it is important to consider that due to the finite nature of
magnetic samples in real life, skyrmions will have a weaker stability since
they can be destroyed through the boundaries and there is no topological
protection. It may be possible to overcome this through geometry, material and
device design.

All data from this study, used to create the figures, can be reproduced from a
repository in Ref.~\onlinecite{Cortes2016b}.

\section*{Methods}
\label{sec:methods}

\subsection*{Material specifications}

Our main study is focused on thin cobalt nanotracks with interfacial DMI. In
these systems, skyrmions are stabilised with the help of an anisotropy
perpendicular to the disk or stripe plane. Within a discrete spin model, the
Hamiltonian for a cobalt system of $P$ atomic sites is described as

\begin{equation}
\begin{split}
    H = & - \sum_{\langle i,j \rangle}^{P} J_{ij} \mathbf{s}_i \cdot \mathbf{s}_j
    + \sum_{\langle i,j \rangle}^{P} \mathbf{D}_{ij} \cdot \left[ \mathbf{s}_i \times \mathbf{s}_j \right] \\
    & - \sum_{i}^{P} \mathcal{K}_{\text{u}}
      \left( \mathbf{s}_i \cdot \hat{\mathbf{z}} \right)^{2} \\
      & + \frac{\mu_{0} \mu_{\text{s}}^{2}}{4 \pi} \sum^{P}_{\langle i,j \rangle}
      \left[ \frac{\mathbf{s}_{i} \cdot \mathbf{s}_{j}}{r_{ij}^{3}}
          - \frac{3 \left( \mathbf{s}_{i} \cdot \hat{\mathbf{r}}_{ij} \right)
              \left( \mathbf{s}_{j} \cdot \hat{\mathbf{r}}_{ij} \right)
               }{ r_{ij}^{5} }
        \right]
  \end{split}
\end{equation}

\noindent where the normalised vector $\mathbf{s}_{i}$ is the spin direction at
the $i$th site, $J_{ij}$ and $\mathbf{D}_{ij}$ are the atomistic exchange and
DMI tensors between the spin at the $i\text{-th}$ site with the $j\text{-th}$
nearest neighbour respectively, which have integrated the $S^{2}$ factor, $S$
being the total average spin per lattice site. For an interfacial DMI, we can
write the Dzyaloshinskii vector as~\cite{Yang2015a}
$\mathbf{D}_{ij}=D\hat{\mathbf{r}}_{ij}\times\hat{\mathbf{z}}$. Moreover,
$\mathcal{K}_{\text{u}}$ is the anisotropy constant per lattice site,
$\mu_{\text{s}}=g\mu_{\text{B}}S$ is the magnetic moment with $g$ as the
Land\'e $g$-factor and $\mu_{\text{B}}$ the Bohr magneton, and
$r_{ij}=\left|\mathbf{r}_{i}-\mathbf{r}_{j}\right|$ is the distance between two
lattices sites, with $\mathbf{r}_{i}$ the position vector of the $i$th spin.
The summations with the restriction $\langle i,j \rangle$ means counting pair
of spins only once.

In the continuum, the DMI can be theoretically described by the energy density
in terms of two Lifshitz invariants. Following the formalism of Rohart and
Thiaville~\cite{Rohart2013} the invariants are defined with an opposite
chirality~\cite{Sampaio2013}:

\begin{equation}
    w_{\text{DM}} = -D_{\text{c}} \left( \mathcal{L}^{(x)}_{xz} + \mathcal{L}^{(y)}_{yz} \right).
\end{equation}

\noindent where $D_{\text{c}}>0$ is the DMI constant. Therefore, for a
hexagonal lattice, we use the following relations to convert the micromagnetic
parameters into atomistic values

\begin{equation}
    A = \frac{\sqrt{3}J_{ij}}{2 a_{z}}
,\, D_{\text{c}} = \frac{\sqrt{3} D}{a a_z}
,\, M_{\text{s}} = \frac{g \mu_B S}{\frac{\sqrt{3}}{2}a^2 a_z}
,\, K_{\text{u}} = \frac{\mathcal{K}_{\text{u}}}{\frac{\sqrt{3}}{2}a^2 a_z}
\label{eq:micro-to-atomistic}
\end{equation}

\noindent where $a$ is the lattice constant in the atomic layer plane, $a_{z}$
is the interlayer spacing, $A$ is the exchange constant, $M_{\text{s}}$ the
saturation magnetisation and $K_{\text{u}}$ the anisotropy constant. Using
these formulas, we notice that the skyrmion size depends on the dipolar
interactions and our atomistic simulations show larger skyrmions than in the
continuum model. The standard derivation of the demagnetising field from a
discrete model in micromagnetics is based on Brown's approximations under
Lorentz assumptions where, for linearly changing magnetisation fields (or
symmetrical square lattices), there is an anisotropic term that is usually not
taken into account since it averages to zero~\cite{BrownJr.1963,Aharoni2000}.
The nonlinearity of the skyrmion structure probably falls outside these
approximations, causing the phenomenon we have observed. This issue has been
previously mentioned in the literature~\cite{Adam1972,Jourdan2008a}.  To make
the atomistic results comparable to the micromagnetic ones~\cite{Sampaio2013},
we assume atomistic lattice distances with equal magnitudes along the plane and
layer thickness, \ie{} $a=a_{z}=2.5\,\text{\AA}$ (rather than
$a_{z}=4\,\text{\AA}$), obtaining good agreement in skyrmion dimensions for the
discrete model. Accordingly, based on the micromagnetic parameters specified on
Ref.~\onlinecite{Sampaio2013} and using equations~\ref{eq:micro-to-atomistic},
the atomistic magnetic parameters are $\mu_{\text{s}}=0.846\,\mu_{\text{B}}$,
$J_{ij}=27.026\,\text{meV}$ and $\mathcal{K}_{\text{u}}=0.0676\,\text{meV}$. To
match the track system of 80 by 40~nm, we specified a lattice of $320\times185$
atoms.

\subsection*{Nudged Elastic Band Method}

Continuing the discussion of the NEBM section, the evolution of the energy
bands are made using a first order differential equation where every image is
evolved with a fictional time $\tau$.  In Cartesian coordinates we use a LLG
kind of equation, based on the work of Suess et al.~\cite{Suess2007}:

\begin{equation}
    \frac{\partial \mathbf{Y}_i}{\partial \tau} = -\mathbf{Y}_{i} \times
    \mathbf{Y}_{i} \times \mathbf{G}_{i} + c \sqrt{ \left( \frac{\partial \mathbf{Y}_{i}}{\partial \tau} \right)^{2} } %
			\left( 1 - \mathbf{Y}_{i}^{2} \right) \mathbf{Y}_{i}
\label{eq:LLG-NEB-cartesian}
\end{equation}

In equation \ref{eq:LLG-NEB-cartesian}, the last term is to constrain the
length of the spins using a suitable factor $c$.  The vector
$\mathbf{G}(\mathbf{Y}_{i}) = \mathbf{G}_{i}$ is defined as a force (for every
image) perpendicular to the band plus a spring force parallel to the band (see
Ref.~\onlinecite{Dittrich2002,Henkelman2000})

\begin{equation}
    \mathbf{G}_{i} =  - \boldsymbol{\nabla}_{\boldsymbol{\mu}} E(\mathbf{Y}_{i})|_{\perp} +
                 \mathbf{F}(\mathbf{Y}_{i})|_{\parallel}
\end{equation}

\noindent The gradient in this definition is with respect to the spin moment
$\boldsymbol{\mu}=\mu_{\text{s}}\mathbf{s}$, which gives the number of degrees
of freedom, hence we take advantage of the effective field definition when
evaluating the gradient:
$\boldsymbol{\nabla}_{\boldsymbol{\mu}}E(\mathbf{Y}_{i}) =
\mu_{\text{s}}^{-1}\partial E / \partial\mathbf{s}=-\mathbf{H}_{\text{eff}}$.  The
parallel component means following the direction of a tangent vector
$\mathbf{t}_{i}$ of an image $\mathbf{Y}_{i}$ which depends on the energy of
its neighbours~\cite{Henkelman2000,Dittrich2002}. In the Climbing Image
technique, we redefine the $\mathbf{G}$ vector for a single image (the climbing
image) of the energy band, which is usually close to a saddle point,
as~\cite{Henkelman2000a}

\begin{equation}
    \mathbf{G}^{\text{CI}}_{i} =  - \boldsymbol{\nabla}_{\boldsymbol{\mu}} E(\mathbf{Y}_{i})|_{\perp} +
                 \boldsymbol{\nabla}_{\boldsymbol{\mu}} E(\mathbf{Y}_{i})|_{\parallel}
\end{equation}

When evolving the system using equation \ref{eq:LLG-NEB-cartesian}, the
tangents $\mathbf{t}$ and effective force $\mathbf{G}$ have been projected into
the tangent space to avoid misbehaviour of the band when the forces overlap, as
specified in Ref.~\onlinecite{Bessarab2015}.

The spring force is defined using the images distance

\begin{equation}
    \mathbf{F}(\mathbf{Y}_{i})|_{\parallel}=k\left(|\mathbf{Y}_{i+1}-\mathbf{Y}_{i}|-
    |\mathbf{Y}_{i}-\mathbf{Y}_{i-1}|\right)\mathbf{t}_{i}.
\label{eq:Spring-force}
\end{equation}

\noindent Correspondingly, the images distances are defined by a Geodesic
length using Vincenty's formulae~\cite{Bessarab2015}.

Most of the figures have the abscissa defined as as the distance from the first
image of the band $\mathbf{Y}_{0}$. This simply means summing up the distances
from neighbouring images, \ie{} the $i$th image in the band will be at a
distance

\begin{equation}
    d = \sum_{j=0}^{i-1} \left| \mathbf{Y}_{j+1} - \mathbf{Y}_{j} \right|
\end{equation}

\noindent from the first extreme of the band, which we measure in arbitrary
units.

We describe the magnetisation using Cartesian coordinates, although any
suitable coordinate system could also be used~\cite{Bessarab2015}. Our approach
to determine the minimum energy band is to set an initial energy band using a
linear interpolation of the spin angles between the fixed equilibrium images
$\mathbf{Y}_0$ and $\mathbf{Y}_{N-1}$ in spherical coordinates, and then to
evolve the band in the chosen Cartesian coordinates to find the minimum energy
transition path between the fixed images. The advantages of using a Cartesian
description of the spins is that the energy band is better defined when the
spins directions are close to the poles but it is still necessary to constrain
their length, which is fixed at zero temperature.  While we discuss the initial
energy band, we note that is also possible to manually specify an initial guess
for the transition, usually by taking one or more images from a known path, for
example the intermediate states when applying a spin polarised current between
two equilibrium states.

We note that with the NEBM one has to hope that chosen initial paths capture
the physically relevant transition paths with the lowest energy barriers when
minimised, but this cannot be proven. There are certainly many other transition
paths with higher energy barriers but they are not important for the system
stability. If we had missed such a transition, this would mean that there is a
transition path with an energy barrier even smaller than the reversal via the
boundary that we have identified.

Using the standard definition for the magnetisation in spherical angles,
$\mathbf{s}=(\sin\theta \cos\psi, \sin\theta \sin\psi, \cos\theta)$, the linear
interpolation initial state of the magnetisation is obtained through the angles
of corresponding spins, between two different images, say $\mathbf{Y}_{i}$ and
$\mathbf{Y}_{k}$ with $i<k$ (we usually use the extreme images, hence $i=0$ and
$k=N$).  Thus, for every spin $j\in\{0, \ldots, P-1\}$ of the system, if we
perform $n$ interpolations, the interpolated angles $(\theta_{j}^{(l)},
\psi_{j}^{(l)})$ of the image $\mathbf{Y}_{l}$, $l \in \{ i,\ldots,k \}$, are

\begin{equation}
    \begin{split}
        \theta_{j}^{(l)} = & \theta^{(i)}_{j}
        + \frac{l}{n + 1} \left[ \theta^{(k)}_{j} - \theta^{(i)}_{j} \right] \\
        \psi^{(l)}_{j} = & \psi^{(i)}_{j}
        + \frac{l}{n + 1} \left[ \psi^{(k)}_{j} - \psi^{(i)}_{j} \right]
    \end{split}
\end{equation}

Regarding the spring force, Dittrich et al.~\cite{Dittrich2002} stated that
their results did not require its application when using a variable order and
time step method.  In our study, the spring force has an influence in the
results, affecting the number of iterations necessary for the algorithm
converge and to achieve a better equispaced band. An estimation of the spring
constant $k$ in equation \ref{eq:Spring-force} is difficult to compute, because
it depends on many factors of the NEBM, such as the size of the system, number
of spins, interactions involved or the coordinates system.  We performed a
series of tests to check optimal values for $k$, and it is usually in a range
around $10^{2}$ to $10^{5}$. For larger order of magnitudes, the algorithm
requires significant computation time, specially when using a small criteria
for stopping the algorithm.

We define the convergence of the NEBM as follows. We first calculate the norms
of the difference (in corresponding degrees of freedom) between the energy
bands of the last NEBM step and the previously computed step. Consequently we
scale them by the number of degrees of freedom (spins) per image and finaly
compute the maximum of these norms and divide by the last time discretisation
given by the integrator~\cite{Sundials}. Thus, we say the band converged if
this value if smaller than a specified criteria. We usually use a value about
$10^{-6}$.

\subsection*{Topological charge}

For a hexagonally arranged discrete spins lattice, we use a topological charge
$Q$ defined for discrete lattices, which was proposed by Berg and
L\"{u}scher~\cite{Berg1981} and, recently, applied by Yin et al.~\cite{Yin2016}
to square arrangements. This is based on taking two spherical triangles per
lattice site, where every triangle is defined by three neighbouring spins,
covering the whole lattice area. This mapping counts the number of times the
spin directions cover a unit sphere. For details, see the Topological charge
section in the Supplementary Information.

\subsection*{Computational simulations}

The NEBM atomistic simulations were performed with Fidimag~\cite{Fidimag2}, a
finite differences code written in Python and C, that uses
Sundials~\cite{Sundials} for integrating the dynamical equations. In addition,
Matplotlib~\cite{Matplotlib}, IPython~\cite{iPython} and the Jupyter
notebook~\cite{Jupyter2016} were used for data analysis, and
Mayavi~\cite{Ramachandran2011} and Povray~\cite{Povray} for the figures.


\section*{Acknowledgements}

We acknowledge financial support from CONICYT Chilean scholarship programme
Becas Chile (72140061), EPSRC's Centres for Doctoral Training in Next
Generation Computational Modelling (EP/L015382/1) and Complex Systems
Simulations (EP/G03690X/1), the EPSRC's Programme grant on Skyrmionics
(EP/N032128/1), the Horizon 2020 European Research Infrastructure project
OpenDreamKit (676541), and the Gordon and Betty Moore Foundation through Grant
GBMF \#4856, by the Alfred P. Sloan Foundation and by the Helmsley Trust.

\section*{Author contributions}

D.C.-O., W.W. and H.F. devised the study and D.C.-O. carried out the
simulations and generated the data. D.C.-O. and H.F. prepared the manuscript.
D.C.-O., H.F., M.B., R.A.P. and O.H. analysed and discussed the results.
D.C.-O.  and W.W.  implemented the NEBM algorithm into the atomistic
simulations framework software. D.C.-O., W.W., M.B., R.A.P., M.A.-B., R.C.,
M.V., T.K. and H.F.  mantained, developed and tested the software. We thank P.
F. Bessarab for data sharing and helpful discussions regarding the NEBM.

\section*{Competing financial interests}

The authors declare no competing financial interests.



\section*{Supplementary material}

# Thermal stability and topological protection of skyrmions in nanotracks (Supplementary Information)

David Cortés-Ortuño,[1, *] Weiwei Wang,[1, 2] Marijan Beg,[1] Ryan A. Pepper,[1] Marc-Antonio Bisotti,[1] Rebecca Carey,[1] Mark Vousden,[1] Thomas Kluyver,[1] Ondrej Hovorka,[1] and Hans Fangohr[1, 3, †]

[1]*Faculty of Engineering and the Environment, University of Southampton, Southampton, SO17 1BJ, United Kingdom*
[2]*Department of Physics, Ningbo University, Ningbo, 315211, China*
[3]*European XFEL GmbH, Holzkoppel 4, 22869 Schenefeld, Germany*

**TEST MODEL**

Bessarab et al.[S1] proposed a toy model to simulate a skyrmion on a discrete spins lattice in order to test the Nudged Elastic Band Method (NEBM). The model Hamiltonian reads

$$H = -\sum_{\langle i,j \rangle}^{N} J_{ij}\mathbf{s}_i \cdot \mathbf{s}_j - \sum_{\langle i,j \rangle}^{N} D_{ij}\left[\mathbf{s}_i \times \mathbf{s}_j\right] - \sum_{i}^{N} \mu_i \mathbf{B} \cdot \mathbf{s}_i \quad (1)$$

which has the convention $\langle i, j \rangle$ in the sums, indicating a summation over every spin pair without repetition. The magnetic parameters are: an exchange constant of $J = 10\,\text{meV}$, a DMI constant of $D = 66\,\text{meV}$, a magnetic moment of $\mu_s = 2\mu_B$ and an out of plane magnetic field of 25 T. Accordingly, we reproduced the system with our finite differences code and obtained a good agreement using Geodesic distances to distinguish the images and Cartesian coordinates to describe the spins. The result is shown in Supplementary Fig. S1, where an energy barrier of approximately 40 meV, with respect to the skyrmion energy, was found after a reasonable number of NEBM iterations.

This test model can be reproduced using the code from the repository in Ref. S2.

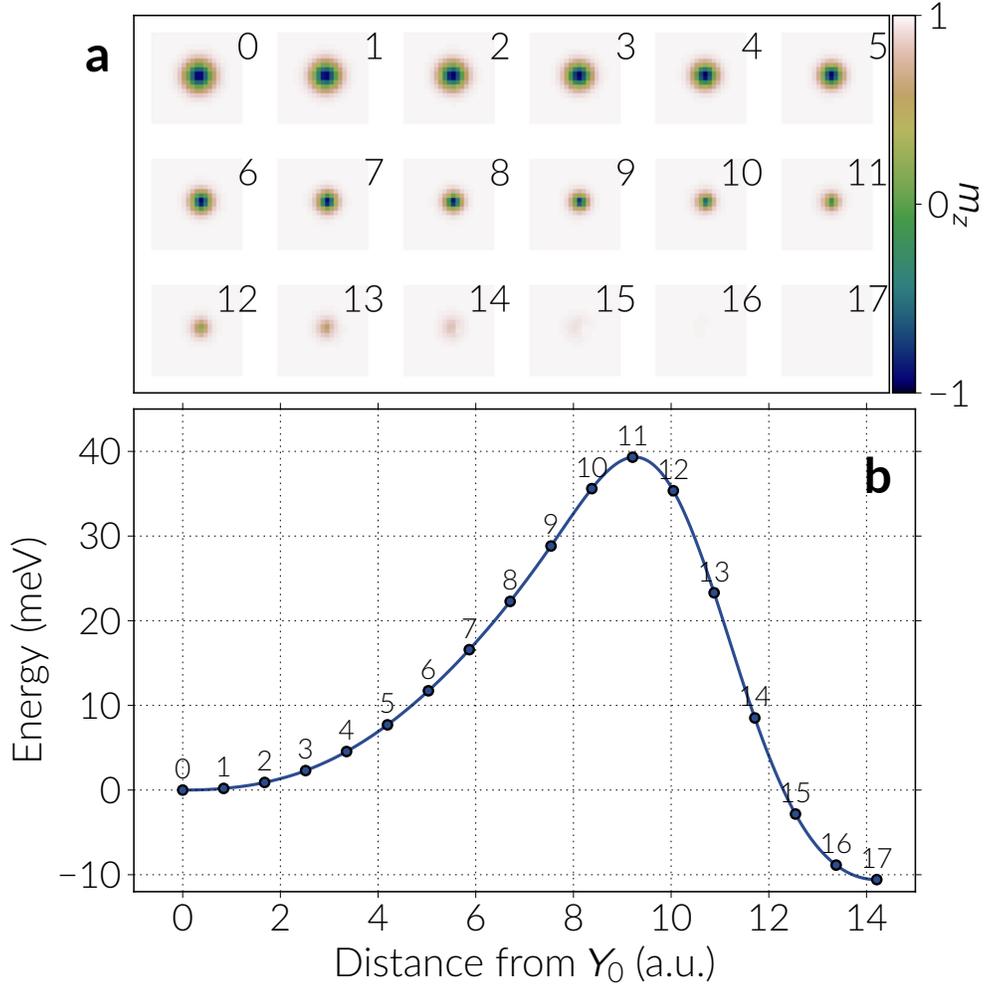

SUPP. FIG. S1. **Energy band of of a skyrmion in a 21x21 spins system**. Spins are arranged in a two dimensional square lattice. Supplementary Figure (b) shows the images of the band depicted in Supplementary Fig. (a), colourised according to the out of plane spin directions magnitude.

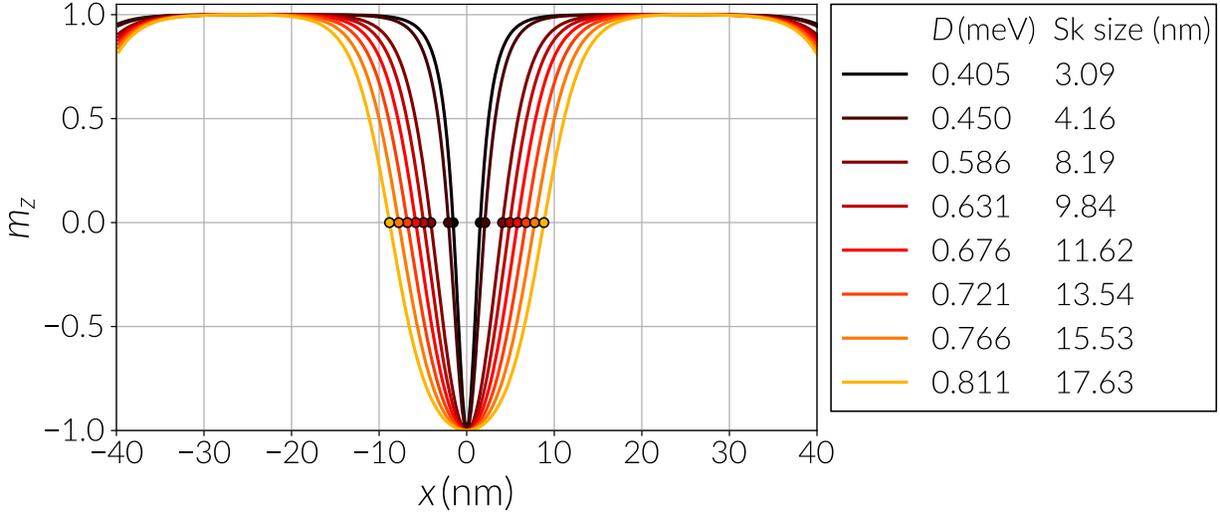

SUPP. FIG. S2. **Skyrmion profile as a function of the DMI magnitude**. For every case we show a continuous line that connects the discrete data points $m_z$ from every lattice site across the nanotrack long side (in the $x$ direction). Small circles indicate points where $m_z = 0$ and using the distance between them we obtain the skyrmion size.

**NANOTRACKS**

In the main text we specified a range of DMI values to analyse the minimum energy transitions of the skyrmions. Specifically, we analysed skyrmions for DMI values from $D = 0.586\,\mathrm{meV}$ to $0.811\,\mathrm{meV}$ in steps of $0.045\,\mathrm{meV}$. We obtained these skyrmions after relaxing them using the Landau-Lifshitz-Gilbert equation. In Supplementary Fig. S2 we show the skyrmion profile for every case, by plotting the out of plane magnetisation component $m_z$ at the middle and across the long edge of the tracks. Using the convention of defining the skyrmion size as the width within the points where $m_z = 0$, we can see that the skyrmion size increases with larger DMI magnitudes. In addition, there is a small tilting at the boundaries which makes a skyrmion on the system to have a magnitude for the total topological charge, slightly smaller than one (or not perfectly zero for the ferromagnetic state).

**Initial states**

Given an initial energy band, the Nudged Elastic Band Method (NEBM) will determine the lowest energy transition path between two equilibrium configurations that is accessible

4

from the initial band. In general, it is difficult to predict an optimal initial guess for the bands, in particular for a system with such a high degree of complexity. In the context of the skyrmion transition, linear interpolations seem a robust starting point and are easily reproducible and applicable to different magnetic systems. A similar technique for initialising the system, where spin directions are interpolated, is using Rodrigues formula[S1].

After exploring the system numerically, we also noticed the major role of the boundaries in the skyrmion annihilation and it was necessary to create an initial path close to this transition, which yields lower energy transition paths. Another way of guessing an initial path would be, for example, to use the dynamics that the system follows when applying spin polarised currents (as Sampaio et al. do in Ref. S3) or magnetic fields (that is not the best choice since the Zeeman energy changes the energy landscape configuration). For the systems of this study, these paths converge to the same results obtained with linear interpolations .

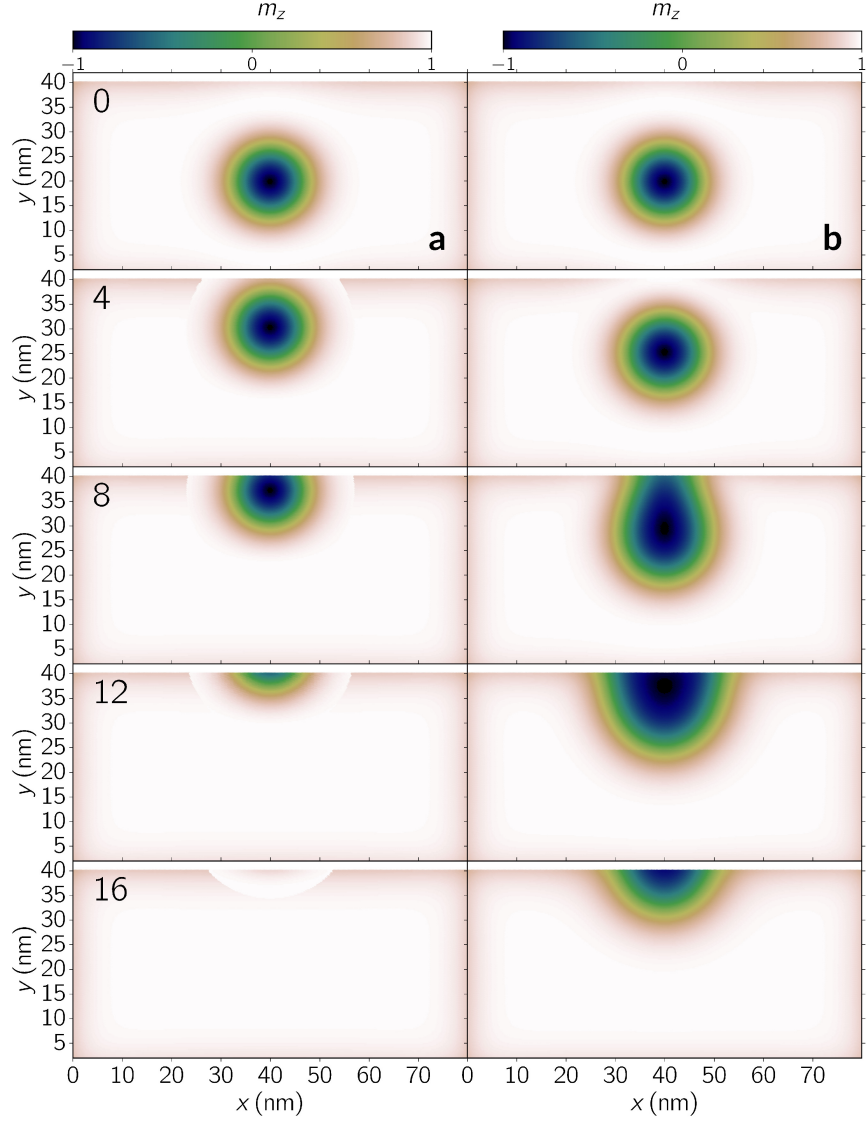

SUPP. FIG. S3. **Initial and final bands for the boundary annihilation energy path**. We use a band of 21 images based on a system with a DMI magnitude of $D = 0.721\,\text{meV}$. Numbers on the top left of every row indicate the image number in the band. (a) Initial state for the NEBM algorithm. (b) Skyrmion annihilation through a boundary obtained after relaxing the band of column (a) with the NEBM.

**Energy band images**

In our study we found three main energy paths for the skyrmion annihilation. For clarity, we provide here more detailed versions of the bands for every transition. Firstly, in Supplementary Fig. S3 we show the skyrmion annihilation through the boundary of the nanotrack, when using a DMI magnitude of $D = 0.721\,\mathrm{meV}$ and a band made of 21 images. In the first column of the figure, *i.e.* Supplementary Fig. S3a, we plot the initial state for the NEBM algorithm, which we manually generate, thus we can see that the skyrmion boundaries are not perfectly defined, specially when approaching the stripe edge. The numbers at the top left of Supplementary Fig. S3 indicate the image number. Consequently, we apply the NEBM to this band and obtain the sequence shown in Supplementary Fig. S3b, where helicoidal structures when the skyrmion is destroyed at the boundary are well defined, and the band has overall reduced its energy. Interestingly, the helicoidal structures are similar to those shown in in Ref. S3, when skyrmions are not successfully generated with spin polarised currents.

Similarly, in Supplementary Fig. S4 we show the skyrmion destruction transition in three different stages, when using a DMI magnitude of $D = 0.721$ meV and a band made of 27 images. In Supplementary Fig. S4a we depict the initial band when using linear interpolations on the spherical angles for the magnetisation field. The numbers on the top left indicate the image number in the band. Consequently, Supplementary Fig. S4b illustrates the sequence after relaxing the initial state with the NEBM, where the final state is the skyrmion destruction by a singularity that resembles a Bloch point, as discussed in the main text. Finally, the last column, shown in Supplementary Fig. S4c, is the band of Supplementary Fig. S4b relaxed with the climbing image NEBM, which we applied to the largest energy point of the band. For this case we clearly see the skyrmion collapse path.

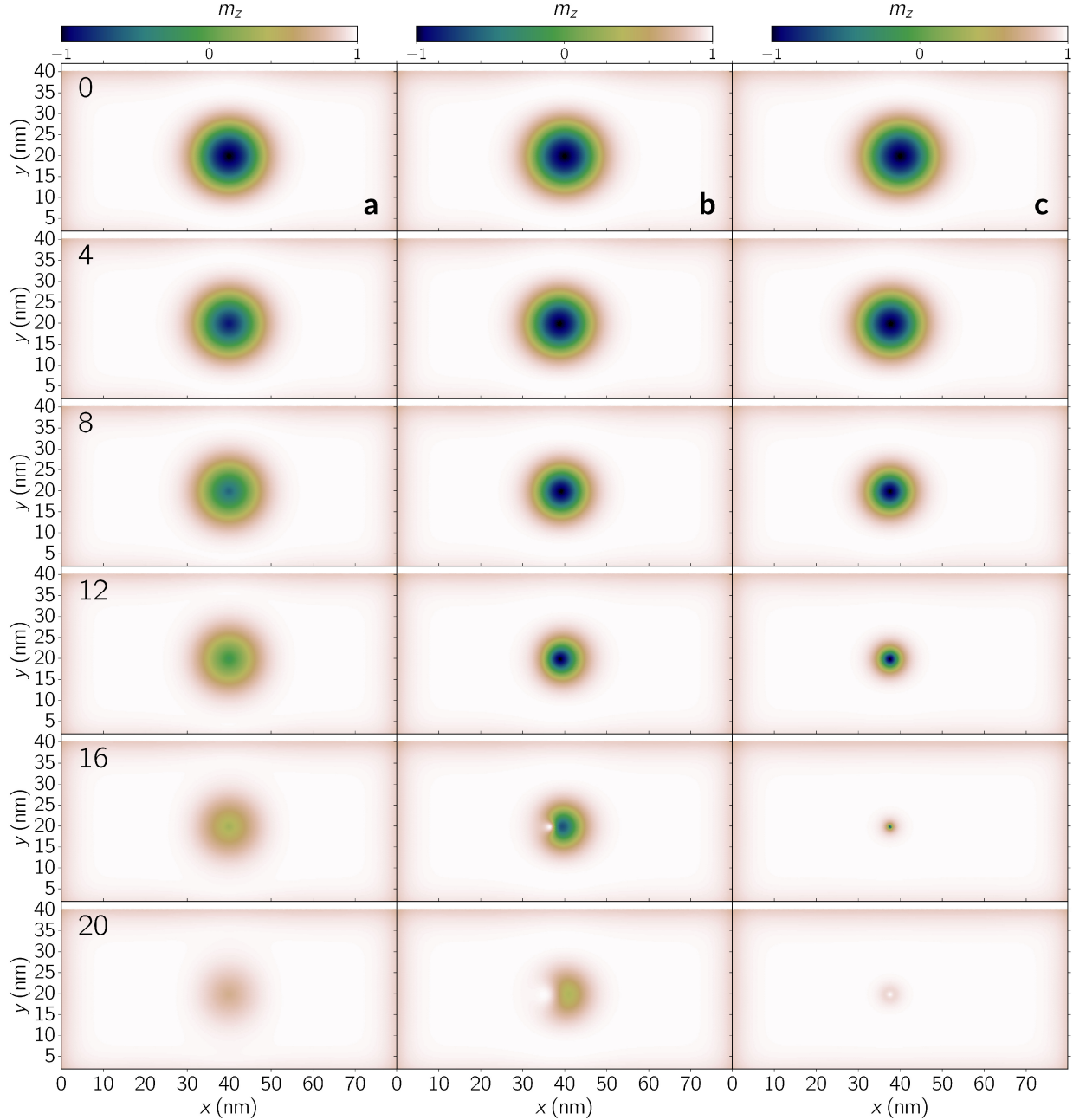

SUPP. FIG. S4. **Skyrmion destruction energy paths in three steps**. We use a band of 27 images based on a system with a DMI magnitude of $D = 0.721\,\mathrm{meV}$. Numbers on the top left of every row indicate the image number in the band. **(a)** Initial state obtained using linear interpolations. **(b)** Skyrmion destruction mediated by a singularity, after relaxing the band of column (a) with the NEBM. **(c)** Skyrmion collapse path, after relaxing the band of column (b) with the climbing image NEBM.

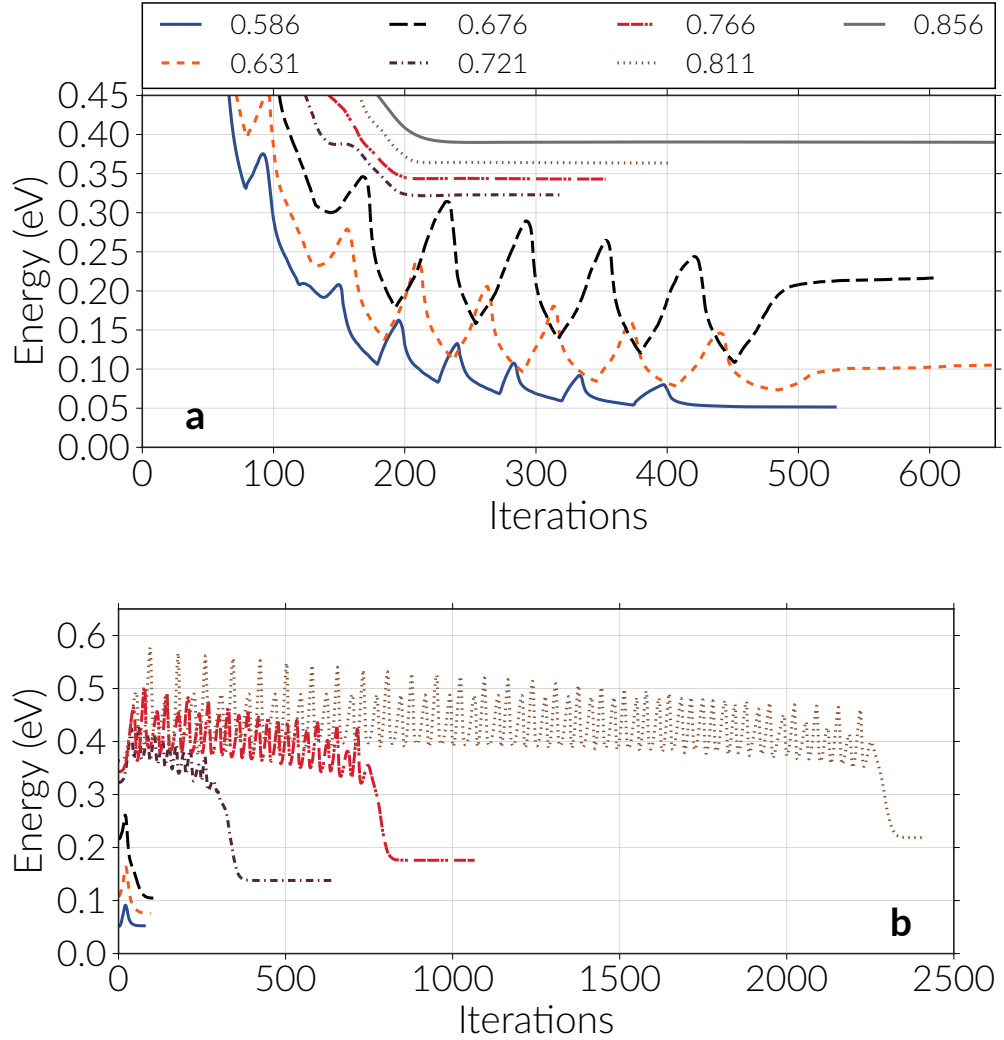

SUPP. FIG. S5. **Evolution of the maximum energy of the bands for the skyrmion destruction processes**. The evolution is computed with respect to the number of iterations of the NEBM relaxation equation (see Methods in the main text) and we show this process for different DMI values. The evolution is calculated using the images with largest energy in the bands. These images can change since they are allowed to move during the iterations. **(a)** Evolution using the NEBM and starting with linear interpolations for the initial bands. **(b)** Evolution using the climbing image NEBM, applied to the bands of (a) after relaxation (last NEBM step).

**Skyrmion destruction evolution**

According to our implementation of the NEBM, we checked how the maximum energy of the band evolves according to the iterations of the algorithm. We compute this magnitude

9

using the image with the largest energy in the band, which can change since images move as the NEBM evolves. In Supplementary Fig. S5 we show this evolution for the case of the skyrmion destruction discussed in the main text. In particular, Supplementary Fig. S5a refers to the skyrmion destruction where the bands for $D = 0.721\,\text{meV}$ and above relax to the singularity mediated skyrmion destruction, and for magnitudes of $D = 0.676\,\text{meV}$ and below, the bands relax to the skyrmion collapse. We can observe that for the smaller DMI values, the maximum energy value oscillates before reaching convergence and this is due to the movement of images around the saddle point (where the last spins defining the skyrmion center flip) whose energy fluctuate dramatically when crossing this region.

Moreover, Supplementary Fig. S5b depicts the evolution of the climbing image NEBM technique applied to the relaxed bands of Supplementary Fig. S5a and all the bands relaxed towards the skyrmion collapse. Bands with $D = 0.721\,\text{meV}$ and above take a long time to relax, specially when the DMI gets stronger. As we specified in the main text, we believe this is due to the rough shape of the energy landscape around the saddle point.

### ENERGY BARRIERS FOR WEAK DMI

We have analysed the energy barriers for DMI magnitudes in the range 0.586 to 0.811 meV. From Supplementary Fig. 5 in the manuscript, the tendency is that the skyrmion collapse curve decreases almost linearly in this range and extrapolating the curve it seems that at some small $D$ value it will cross the boundary annihilation curve. We then computed the values of the energy barriers for DMI magnitudes weaker than 0.586 meV. Decreasing in steps of 0.045 meV, we could not stabilise a skyrmion for $D$ smaller than 0.405 meV, probably due to the strong exchange and anisotropy which are kept fixed. For this small DMI magnitude a skyrmion is only 3.09 nm wide.

We show in Supplementary Fig. S6 the energy barriers computed for the whole range of DMI values in the main study and down to 0.405 meV. In this plot, we show both the energy barriers obtained using the image with the largest image in a band, which we call NEBM curve, and the energy barriers obtained after performing a polynomial interpolation on the bands, named Interpolation curve. In general, the interpolation gives optimal results for the boundary annihilation, which can be seen from Supplementary Fig. 4 in the manuscript. For the skyrmion collapse the polynomial approximation appears to agree well with the largest

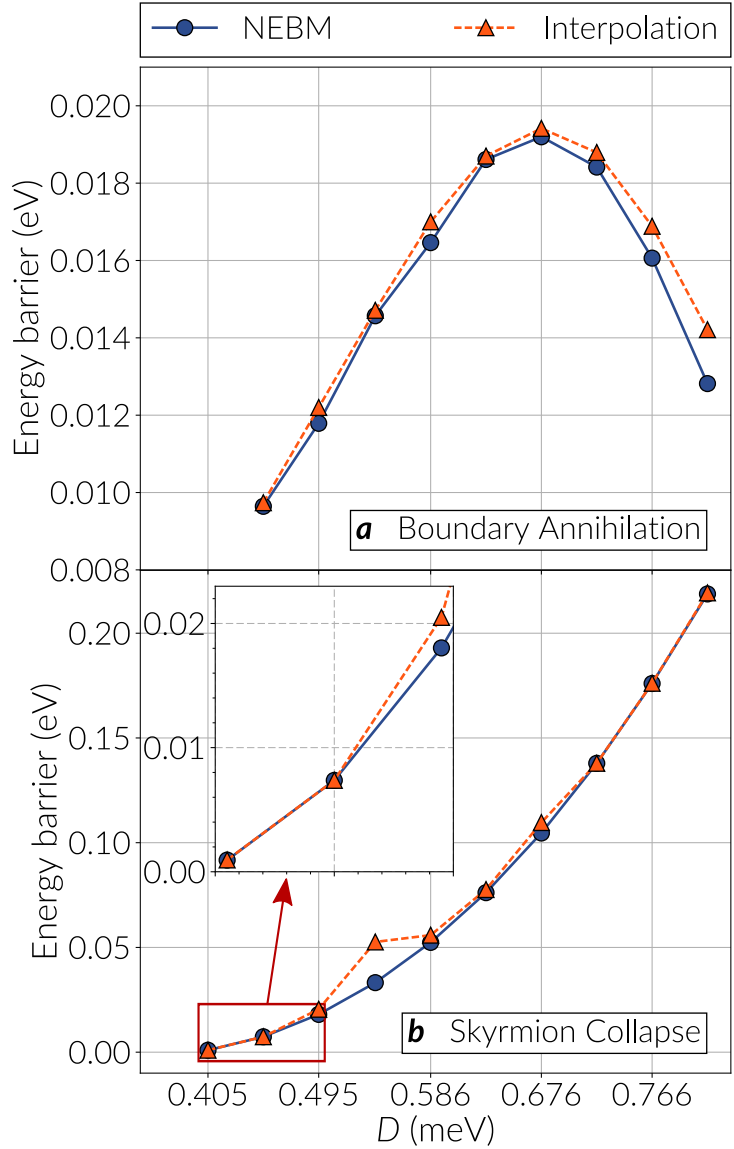

SUPP. FIG. S6. **Calculation of the energy barriers for an extended range of DMI magnitudes**. The NEBM curves show the energy barriers using the image with largest energy in the energy bands and the Interpolation curve are barriers computed when using a cubic polynomial interpolation on the bands. The energy barriers are calculated with respect to the skyrmion energy. The skyrmion cannot be stabilised below 0.405 meV. For $D = 0.541$ meV the polynomial interpolation overestimates the value of the energy barrier. The inset shows the skyrmion collapse energy barriers for the three smallest DMI magnitudes.

point of the band in most cases, but sometimes it can slightly overestimate the energy barrier when there is no good resolution around the saddle point, which is given by the reversal of the spins at the skyrmion nucleus. In particular, this happens for $D = 0.541$ meV, which is shown in Supplementary Fig. S6b. Nevertheless, the quadratic tendency is clear when using the points from the NEBM curve. Interestingly, when comparing the inset of Supplementary Fig. S6b with Supplementary Fig. S6a at $D = 0.450$ meV, where the skyrmion is only 4.16 nm wide, the energy barrier of the skyrmion collapse is slightly smaller. Decreasing the DMI even further we could not obtain a clear value for the boundary annihilation, however in this range the skyrmion collapse barrier is smaller than 0.002 meV.

## TOPOLOGICAL CHARGE

The topological charge gives us an idea about the contribution of the spin directions to the mapping of the magnetisation field into a unit sphere. This helps us to characterise non trivial magnetic structures, such as the Bloch point like structure we observed in our simulations. From our results, a full mapping of the spins around this topological singularity would not be optimal, since the full structure usually lies between two images around a saddle point. Nevertheless, it is still possible to observe large variations of the topological charge density when the singularity starts to appear, allowing us to distinguish it in the spin field.

In micromagnetics, instead of calculating a topological charge, it is standard to compute the so called skyrmion number, which requires the calculation of the derivatives of the magnetization field in a two dimensional plane. For a discrete spins lattice we use the topological charge definition rather than discretising the derivatives along the 2D-lattice directions (since, for the sake of numerical calculations, we have to anyways discretise the continuum micromagnetic mesh), as Rohart et al. mention in S4. As we specified in the main text, we base our calculation on Ref. S5 and S6, where we have extended the definition to hexagonally arranged spin lattices. This calculation is based on computing the spherical angle spanned by triplets of neighbouring spins per every lattice. Accordingly, the total topological charge is defined as

$$Q = \sum_\mu q_\mu = \sum_\mu \frac{1}{4\pi} \left( \Omega_{\mu,1,2} + \Omega_{\mu,3,4} \right) \qquad (2)$$

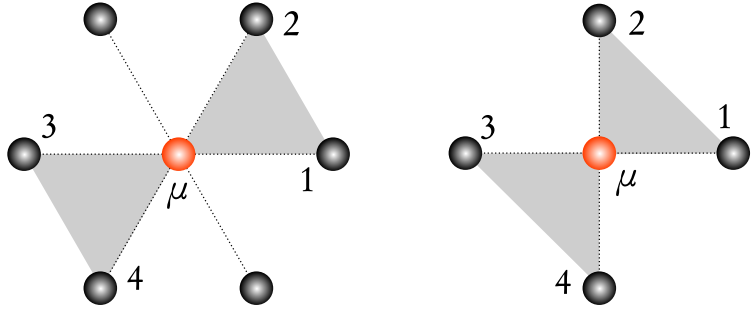

SUPP. FIG. S7. **Triangles of neighbouring spins per lattice site**. Sites are labelled as $\mu$ to define a topological charge or skyrmion number for discrete hexagonal (left image) and square (right image) lattices.

The spherical angle at the $\mu$-th site $\Omega_{\mu,1,2}$, is defined as with respect to the neighbouring spins $\mathbf{s}_1$, $\mathbf{s}_2$ as

$$\exp\left(\frac{1}{2}i\Omega_\mu\right) = \frac{1}{\rho}\left[1 + \mathbf{s}_\mu \cdot \mathbf{s}_1 + \mathbf{s}_1 \cdot \mathbf{s}_2 + \mathbf{s}_2 \cdot \mathbf{s}_\mu + i\mathbf{s}_\mu \cdot (\mathbf{s}_1 \cdot \mathbf{s}_2)\right] \quad (3)$$

with

$$\rho = \left[2\left(1 + \mathbf{s}_\mu \cdot \mathbf{s}_1\right)\left(1 + \mathbf{s}_1 \cdot \mathbf{s}_2\right)\left(1 + \mathbf{s}_2 \cdot \mathbf{s}_\mu\right)\right] \quad (4)$$

Similarly for the $\mu$-th site $\Omega_{\mu,3,4}$. For a square and hexagonal arrangement, the triangles specified per lattice site are depicted in Supplementary Fig. S7 (it would be equivalent to use the opposite triangles that also cover a unit cell). It is worth noticing that Berg and Lüscher[S5] originally introduced a sign for the exponential in equation 3 but we assume the convention that triangles of spins are taken in the counter-clock wise direction, as Yin et al. do in Ref. S6. Moreover, this angle is not completely defined for the special configurations where $\mathbf{s}_\mu \cdot (\mathbf{s}_1 \cdot \mathbf{s}_2) = 0$ and $1 + \mathbf{s}_\mu \cdot \mathbf{s}_1 + \mathbf{s}_1 \cdot \mathbf{s}_2 + \mathbf{s}_2 \cdot \mathbf{s}_\mu$. For these cases, the exponential falls in a branch cut in the complex plane[S5,S6], which implies a change in the topological charge that can be related to the condition where a skyrmion is created or collapses.

The definition of the density charge only considers two triangles around a lattice site, missing the contribution from the other neighbours, but this is necessary to avoid double counting of areas when summing up all the triangles contributions. For a more complete picture of the topological charge per lattice site, we could also take into account the rest of the nearest neighbours.

13

**SUPPLEMENTARY VIDEOS**

SUPPLEMENTARY VIDEO S1. **Relaxation of an energy band for the skyrmion collapse**. The animation shows the evolution of the NEBM applied to a nanotrack with $D = 0.676$ meV. The initial state at the 0th step is a linear interpolation between the skyrmion (image at the left extreme of the band) and the ferromagnetic (right extreme) states. The band at the final steps represents the skyrmion collapse transition, where the image with the largest energy is the saddle point. Around this image, the corresponding magnetic configurations characterise the reversion of the spins at the skyrmion core.

SUPPLEMENTARY VIDEO S2. **Relaxation of an energy band for the singularity mediated skyrmion destruction**. The animation shows the evolution of the NEBM applied to a nanotrack with $D = 0.721$ meV. The initial state at the 0th step is a linear interpolation between the skyrmion (image at the left extreme of the band) and the ferromagnetic (right extreme) states. The band at the final steps represents the skyrmion destruction via a Bloch point like singularity. The image with the largest energy is the saddle point and around this image, the corresponding magnetic configurations characterise the emergence of the magnetic singularity.

SUPPLEMENTARY VIDEO S3. **Climbing image NEBM applied to the skyrmion destruction by a singularity**. The animation shows the evolution of the climbing image technique applied to the image with largest energy (saddle point) in the band from the last step of Video 2, which is the skyrmion annihilation mediated by a singularity. This system has DMI of $D = 0.721$ meV. In the video, the climbing image is labeled during the algorithm evolution. At the end of the relaxation, the energy band converges to an energy band representing a skyrmion collapse, which has a smaller energy barrier.

SUPPLEMENTARY VIDEO S4. **Annihilation of a skyrmion through a boundary**. This animation shows the first images of an energy band representing the skyrmion annihilation mediated by a boundary. The sequence is restricted to a region around a boundary of the nanotrack where the skyrmion destruction occurs. The animation was obtained from

the NEBM simulation applied to a nanotrack with $D = 0.721\,\text{meV}$ and the top left numbers are the image numbers in the band.

SUPPLEMENTARY VIDEO S5. **Skyrmion collapse**. This animation shows the first images of an energy band representing the skyrmion collapse transition. The sequence is restricted to a region at the centre of the nanotrack where the skyrmion destruction occurs and spins are only shown for some lattice sites. The animation was obtained from the climbing image NEBM simulation applied to a nanotrack with $D = 0.721\,\text{meV}$ and the top left numbers are the image numbers in the band.

SUPPLEMENTARY VIDEO S6. **Skyrmion destruction mediated by a singularity**. This animation shows the first images of an energy band representing the skyrmion annihilation by a Bloch point like singularity. The sequence is restricted to a region at the centre of the nanotrack, where the skyrmion destruction occurs, and spins are only shown for some lattice sites. The animation was obtained from the NEBM simulation applied to a nanotrack with $D = 0.721\,\text{meV}$ and the top left numbers are the image numbers in the band.

---



\end{document}